\def\ha{{H$\alpha$}}
\def\hb{{H$\beta$}}

\def\VEL{\:{\rm km\:s^{-1}}}

\def\OiiiL{[\ion{O}{3}] $\lambda 5007$}
\def\SiiL{[\ion{S}{2}] $\lambda\lambda 6716, 6731$}
\def\NiiL{[\ion{N}{2}] $\lambda\lambda 6548, 6583$}

\def\FeiiL{[\ion{Fe}{2}] 1.644 $\mu$m}

\def\sii{[\ion{S}{2}]}
\def\nii{[\ion{N}{2}]}

\def\oiii{[\ion{O}{3}]}
\def\fe2{[\ion{Fe}{2}]}

\def\hii{\ion{H}{2}}

\documentclass[12pt,preprint]{aastex}

\begin{document}


\newcommand{\MSOL}{\mbox{$\:M_{\sun}$}}  

\newcommand{\EXPN}[2]{\mbox{$#1\times 10^{#2}$}}
\newcommand{\EXPU}[3]{\mbox{\rm $#1 \times 10^{#2} \rm\:#3$}}  
\newcommand{\POW}[2]{\mbox{$\rm10^{#1}\rm\:#2$}}
\newcommand{\SING}[2]{#1$\thinspace \lambda $#2}
\newcommand{\MULT}[2]{#1$\thinspace \lambda \lambda $#2}
\newcommand{\CHINU}{\mbox{$\chi_{\nu}^2$}}
\newcommand{\vsini}{\mbox{$v\:\sin{(i)}$}}
\newcommand{\LSOL}{\mbox{$\:L_{\sun}$}}

\newcommand{\fuse}{{\it FUSE}}
\newcommand{\hst}{{\it HST}}
\newcommand{\iue}{{\it IUE}}
\newcommand{\euve}{{\it EUVE}}
\newcommand{\einstein}{{\it Einstein}}
\newcommand{\rosat}{{\it ROSAT}}
\newcommand{\chandra}{{\it Chandra}}
\newcommand{\xmm}{{\it XMM-Newton}}
\newcommand{\swift}{{\it Swift}}
\newcommand{\asca}{{\it ASCA}}
\newcommand{\galex}{{\it GALEX}}
\newcommand{\cxo}{CXO}


\shorttitle{\hst/WFC3 Observations of M83}
\shortauthors{Blair \etal}


\title{
{An Expanded \hst/WFC3 Survey of M83: Project Overview and Targeted Supernova Remnant Search\footnote{
Based on observations with the NASA/ESA 
{\em Hubble Space Telescope}, obtained at the Space Telescope Science Institute,
which is operated by the Association of Universities for Research
in Astronomy, Inc., under NASA contract NAS5-26555.
}}}

\author{
William P. Blair\altaffilmark{2},
Rupali Chandar\altaffilmark{3}, 
Michael A. Dopita\altaffilmark{4,}\altaffilmark{5}, 
Parviz Ghavamian\altaffilmark{6},
Derek Hammer\altaffilmark{7},
K. D. Kuntz\altaffilmark{2},
Knox S. Long\altaffilmark{7},
Roberto Soria\altaffilmark{8}, 
Bradley C. Whitmore\altaffilmark{7},
P. Frank Winkler\altaffilmark{9}
}
\altaffiltext{2}{The Henry A. Rowland Department of Physics and Astronomy, 
Johns Hopkins University, 3400 N. Charles Street, Baltimore, MD, 21218; 
wpb@pha.jhu.edu, kuntz@pha.jhu.edu}
\altaffiltext{3}{Department of Physics \& Astronomy, University of Toledo, Toledo, OH, 43606; Rupali.Chandar@utoledo.edu}
\altaffiltext{4}{Research School of Astronomy \& Astrophysics, The Australian National University, 
Cotter Road, Weston ACT 2611, Australia; Michael.Dopita@anu.edu.au}
\altaffiltext{5}{Astronomy Department, King Abdulaziz University, P. O. Box 80203, Jeddah, Saudi Arabia.}
\altaffiltext{6}{Department of Physics, Astronomy, \& Geosciences, Towson University, Towson, MD, 21252; pghavamian@towson.edu}
\altaffiltext{7}{Space Telescope Science Institute, 3700 San Martin Drive, Baltimore, MD, 21218;  long@stsci.edu, hammer@stsci.edu, whitmore@stsci.edu}
\altaffiltext{8}{Curtin Institute of Radio Astronomy, Curtin University, 1 Turner Avenue, Bentley WA6102, Australia; roberto.soria@icrar.org}
\altaffiltext{9}{Department of Physics, Middlebury College, Middlebury, VT, 05753; winkler@middlebury.edu}

\begin{abstract}
We present an optical/NIR imaging survey of the face-on spiral galaxy M83, using data from the {\it Hubble Space Telescope} Wide Field Camera 3 (WFC3).  Seven fields are used to cover a large fraction of the inner disk, with observations in nine broadband and narrowband filters.  In conjunction with a deep \chandra\ survey and other new radio and optical ground-based work, these data enable a broad range of science projects to be pursued.   We provide an overview of the WFC3 data and processing and then delve into one topic, the population of young supernova remnants.  We used a search method targeted toward soft X-ray sources to identify 26 new supernova remnants.  Many compact emission nebulae detected in \FeiiL\ align with known remnants and this diagnostic has also been used to identify many new remnants, some of which are hard to find with optical images.  We include 37 previously identified supernova remnants that the data reveal to be $<$0\farcs5 in angular size and thus are difficult to characterize from ground-based data.  The emission line ratios seen in most of these objects are consistent with shocks in dense interstellar material rather than showing evidence of ejecta.  We suggest that the overall high elemental abundances in combination with high interstellar medium pressures in M83 are responsible for this result.  Future papers will expand on different aspects of the  these data including a more comprehensive analysis of the overall supernova remnant population.

\end{abstract}

Subject Headings: galaxies: individual (M83) -- galaxies: ISM  -- supernova remnants

Facilities:  Hubble Space Telescope (WFC3)

\section{Introduction \label{sec_intro}}

M83 (NGC\,5236) is an iconic face-on ($i$ = 24$^{\circ}$) grand-design SAB(s)c spiral galaxy, with a starburst nucleus, active star formation along the arms, and prominent dust lanes \citep{{talbot79}, {elmegreen98}}.   Adopting the Cepheid distance of 4.61 Mpc to M83 \citep{saha06} means that 1\arcsec\,=\,22 pc.  With its proximity and nearly face-on orientation,  M83 has been the subject of numerous studies at wavelengths across the electromagnetic spectrum \citep[to mention a few]{{crost02}, {soria03}, {herrmann08}}, and as each new facility or improved instrument comes on-line, observers return to this galaxy to improve the available data.  This is because  M83 provides an exceptional example for studying the entire cycle of star formation and destruction, and the impacts of this activity on the structure and evolution of the galaxy itself.  Over time, the integrated effects of these ongoing processes reveal themselves in the form of high overall metallicity  \citep{{bres02}, {pily06}, {pily10}} and chemical abundance gradients across the $\sim$10\arcmin\ diameter bright optical disk  \citep{{bres02}, {bres09}}.  GALEX UV imaging and deep H~I surveys show a fainter and much more extended and distorted disk, indicative of past interactions and active feeding of new material to the inner galaxy \citep{{hucht81}, {thilker05}, {bigiel10}}.

One direct indicator of the ongoing activity in M83 is the observed supernova (SN) rate.  To date, M83 has hosted six recorded supernovae (SNe) since 1923, although none since 1983 \citep{{cowan85}, {stockdale06}}, second in number only to NGC\,6946 which has had nine.  
Three of the six have spectroscopically determined types of Ib or II, consistent with them resulting from the core-collapse of massive stars \citep{barbon99}.  
Projecting this observed rate backwards in time, there must have been dozens of core-collapse SNe in M83 within the past millennium, and many more older supernova remnants (SNRs) as well since expectations are that SNRs remain visible for tens of thousands of years.  This is consistent with existing SNR surveys that have identified well over 200 SNR candidates in M83 \citep[see][henceforth B12, and references therein; see also Blair et al. 2013]{blair12}.

We are pursuing a multi-wavelength observational campaign to obtain new optical/IR, X-ray, and radio data for M83 to better understand the various populations of objects and how they interact with each other and with the galaxy as a whole.  As part of this campaign, B12 used IMACS imaging data from the Magellan-I 6.5m telescope  to identify some 225 ISM-dominated SNRs and some additional relatively strong \oiii\ sources that are potential SNRs, expanding on the earlier work of \citet[][henceforth BL04]{blair04}. In addition, \citet{long14} conducted an extensive \chandra\ campaign totaling 729 ks (plus 60 ks from much earlier in the \chandra\ mission) that shows over 400 point sources and extensive diffuse X-ray emission filling the spiral arms and star forming regions.  Radio observations with ATCA (also reported by \cite{long14}) and the Jansky EVLA (Stockdale et al., in preparation) have also been conducted.  Early results include the discovery and characterization of a new ultraluminous X-ray source \citep{soria12}, an improved characterization of the young remnant of SN1957D \citep{long12}, and the discovery of a new microquasar near the nucleus \citep{soria14}.  

In this paper, we provide an overview of the \hst\ portion of our multi-wavelength campaign, which includes imaging in nine broadband and narrow emission-line filters in the optical and near-IR of seven fields in M83, two of which are from the Early Release Science Program (ID 11360; R. O'Connell, PI) and five of which are from a cycle 19 \hst\ General Observer program (12513; W. Blair, PI).  The excellent spatial resolution of WFC3 ($0\farcs0396$ per pixel for UVIS, $0\farcs13$ per pixel for the IR camera) permits accurate stellar photometry in relatively crowded fields and also resolves the emission-line gas at the parsec level.  After this overview, we report initial results from our analysis of the smallest diameter SNR candidates uncovered by \hst/WFC3.  \citet[henceforth D10]{dopita10} performed what in many ways was a precursor to the current SNR investigation, using \hst/WFC3 data from what we will call Field 1 to investigate the SNR population in the M83 nuclear region and an inner spiral arm. They identified 60 SNRs  and candidates within this one WFC3 field, including 20 objects in the complex nuclear region, a possible (but spectroscopically unconfirmed) young ejecta-dominated SNR on the eastern edge of the nucleus, and a tentative optical identification of the counterpart to SN1968L (also deep in the nuclear region).

The next section describes the \hst\ observations and data processing.  In  \S3, we describe the candidate SNRs and their properties, concentrating on the subset for which \hst\ data provide the most leverage.   In \S4 we discuss  the implications of these results, and we summarize our findings in \S5.   We will address various other aspects of these data or specific objects of interest in future papers, including a more comprehensive treatment of the full SNR population and its relation to the underlying stellar component.

\section{Observations and Data Reduction \label{sec_obs}}

The Wide Field Camera 3 (WFC3) was installed in \hst\ in May 2009 during the final servicing mission (SM4).  Details of the instrument and its performance are provided in the WFC3 Instrument Handbook \citep{dressel12} and other documentation available through the Space Telescope Science Institute web site {\tt http://www.stsci.edu/hst}.

\begin{figure}
\plotone{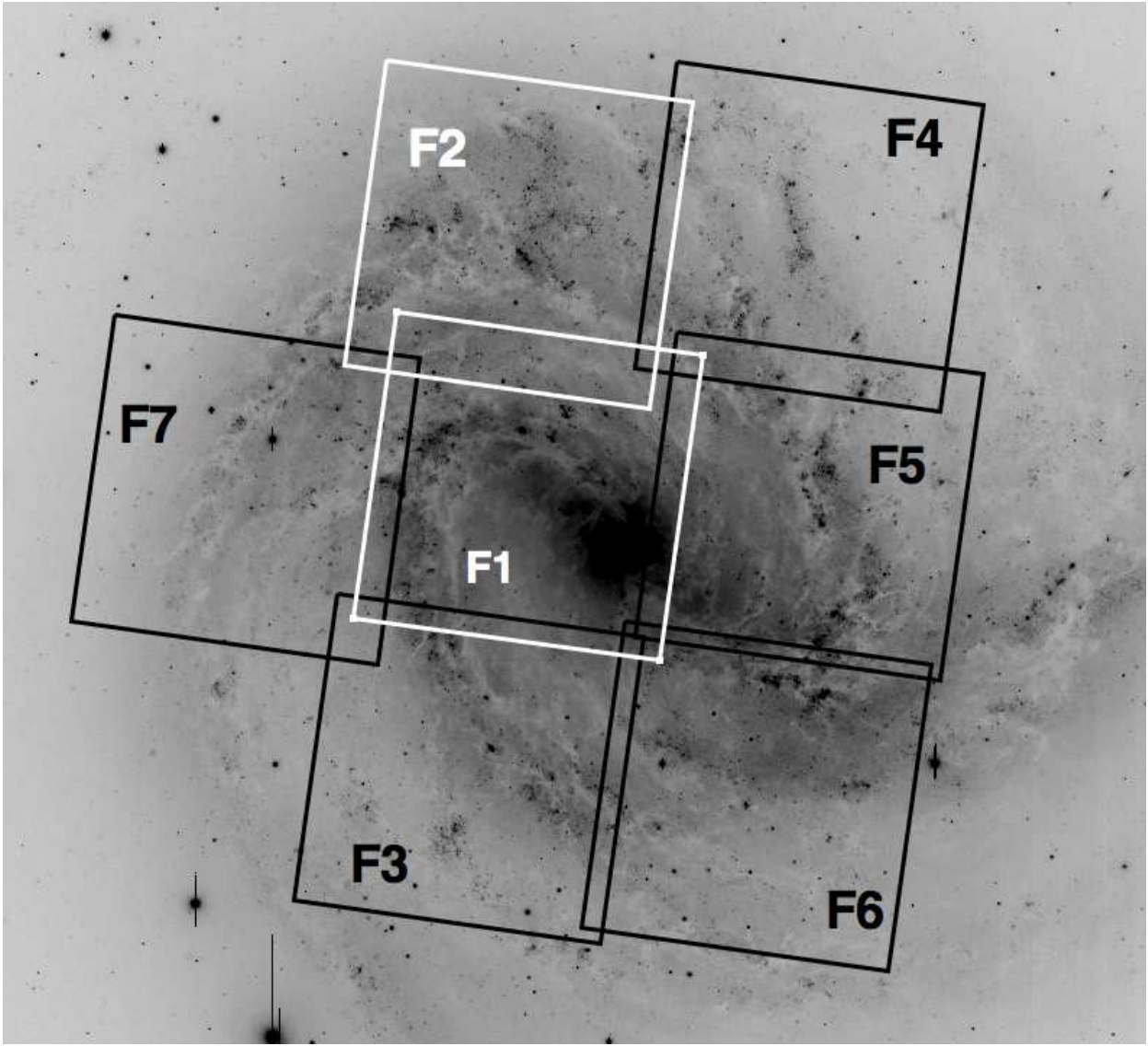}
\caption{The V-band image of M83 from our Magellan data (B12)  is shown with a grid that indicates the approximate locations of the seven WFC3 UVIS fields of view.   For scale, each box is 162\arcsec\ on a side.  The two white boxes show the archival fields from program 11360, while the black boxes show the fields from our cycle 19 program 12513.  The field IDs shown are used throughout this paper.  As with all Figures in this paper, north is up and east is to the left.   \label{fig_overview}}
\end{figure}

The approximate field coverage for the WFC3/UVIS camera fields of view is shown projected onto a ground-based image of M83 in Figure \ref{fig_overview}.  The WFC3/UVIS field of view is $162\arcsec \times 162\arcsec$ but slightly parallelogram in shape, which we have ignored in this approximate representation. The individual fields were overlapped by a small amount to ensure no spatial gaps in coverage for the UVIS fields.  However, the WFC3/IR camera field is smaller, $136\arcsec \times 123\arcsec$.  Rather than overlap the UVIS fields by a larger fraction to ensure full IR coverage,  we chose to accept small gaps in the IR coverage and take the WFC3/IR data at the same nominal field centers as the UVIS data.  

There are some differences in the data due to changes over time in WFC3.  Shortly after WFC3 was installed in \hst, the Early Release Science program of the WFC3 Science Oversight Committee imaged two fields in M83 as part of a larger program studying star formation in various settings.  What we will call Field 1 covered the nuclear region and inner spiral arm to the east of the nucleus, while Field 2 was overlapping Field 1 and extending to the north (see Figure \ref{fig_overview}).   The Field 1 data have been studied the most extensively, with both the SNR population (D10) and the star cluster populations \citep{{kim12}, {whitmore11}, {chandar10}} already published.  Clusters in both fields have been assessed by \citet{bastian12} and by \citet{chandar14}.  Field 2 data were also used in the detailed study of the young remnant of SN1957D \citep{long12}.  The data for these two fields were obtained in August 2009 and March 2010, prior to the onset of charge transfer efficiency (CTE) effects in WFC3/UVIS, which have developed over time \citep[see also {\tt http://www.stsci.edu/hst/wfc3/ins\_performance/CTE/}]{wfc3cte}.  These fields were observed with a more extensive set of filters than has been used for the remainder of the survey and only those filters in common with the new fields are discussed here.

The data for Fields 3 through 7 (as shown in Fig. \ref{fig_overview}) were all obtained as part of Cycle 19 \hst\ program 12513, over the time period from July 2012 through early September 2012.  These observations spanned the time frame over which the adverse impacts of charge transfer efficiency (CTE) were in the process of being characterized and a mitigation strategy developed by STScI \citep{wfc3cte}.  The data for U, B, I, and F657N (\ha) for Fields 4 and 5 were the first of the new data to be obtained, and a quick inspection showed that the background from the galaxy itself was not sufficient over the darkest portions of each field to mitigate CTE.  Hence, the remainder of the program was updated to include post-flash to raise background levels to the approximate levels that mitigate CTE effects.  

The exposure times, FLASH parameter, and other supporting information for each field, including the nine filters used, are summarized in Table \ref{table_obs}.  We note that F555W (V) was used for Field 1, but the remainder of the fields used the medium band F547M (y) filter to avoid any contamination from significant emission lines and thus provide the cleanest possible continuum band.  Photometry in this filter can easily be converted to Johnson V with little loss of accuracy.  As for the emission-line filters, F502N passes \OiiiL, F673N passes both lines of  \SiiL, and F657N is broad enough that it passes both \ha\ and the nearby \NiiL\ lines.  Since previous spectra in M83 \citep{blair04} demonstrate the strength of  \nii\ is substantial in relation to \ha\ (especially in the inner part of the galaxy), this contaminating effect cannot be ignored.
F164N passes the \FeiiL\ line, which turns out to be an interesting and useful new diagnostic for the shock-heated nebulae, which are primarily but not exclusively SNRs (see below).

The exposure time in each filter was split between three roughly equal parts, and the field of view was stepped between sub-exposures to provide for cosmic ray and hot pixel rejection and to cover the chip gap in the cameras.  For UVIS, the field was stepped 1.605\arcsec\ at a pattern orientation of 86.7$^{\circ}$.  For the IR camera, a larger step of 12.99\arcsec\ at a pattern orientation of 65.14$^{\circ}$ was used.  The larger step size for the IR camera was chosen to reduce the loss of field coverage due to its smaller field of view, but of course this comes at the cost of a portion of each field only having coverage in one or two of the three sub-exposures.  While the cosmic ray rejection is less than ideal in these portions of each field, the additional field coverage turned out of be advantageous as a number of objects of interest were covered that otherwise would have been missed.  Even so, some objects of interest ended up in the gaps in IR coverage and were not observed in F164N.

For each field, we constructed drizzle-combined images for all nine
filters with the constraints that the images have the same dimensions and
pixel size ($0\farcs0396$) and be aligned in both World Coordinate System (WCS) and pixel (image)
space. Thus the IR images were resampled as part of this process onto the nominal pixel 
scale of the UVIS camera.  Image alignment was performed using the task {\it tweakreg} and
distortion-corrected drizzled images were constructed using {\it AstroDrizzle}, both of which 
are part of the DrizzlePac software suite \citep[e.g.,][]{Fruchter10,Gonzaga12}. 
The {\it tweakreg} routine finds stars in common 
between the relevant images and uses fitting routines to determine the adjustments 
needed to align to a given reference image. A more detailed description of the procedure 
used to create the final, aligned, drizzled images for each filter is provided in the Appendix. 

After aligning the images for each field individually, we then aligned all of the fields onto a common absolute astrometry scale.  Relative positional shifts in the overlap regions of the various fields were small ($<$0\farcs75) but noticeable.  We used centroids of isolated stars in the overlap regions to align all of the frames to a single grid, and then used stars from the UCAC3 astrometric catalog \citep{zacharias10} and 2MASS, selecting ones with small positional uncertainties and checking visually to eliminate close doubles and a few background galaxies.  The WCS keywords in the FITS file headers were adjusted to place the entire data set on the same accurate absolute scale, consistent at better than 0\farcs1.   All cataloged positions in this paper refer to this absolute astrometric solution.  The data frames were compared to Magellan ground-based imagery of B12, which had also been placed on the same astrometric reference frame, and the alignment was found to be excellent.  

The accurate alignment of all the filter images for each field was an important step of the analysis because it permitted the appropriate continuum images to be combined, scaled, and subtracted from the narrow-band images, thus removing the stars to first order and enhancing the appearance of faint nebular emission.  This is particularly important for finding and studying the smallest diameter emission-line objects which otherwise could easily be mistaken for stars. We used continuum images that bracketed the \oiii, \ha, and \sii\ images, using the central wavelengths of the various filters to balance the relative amounts of each continuum image to use, and using averages of multiple stars in each field to determine relative scaling factors. Because the broadband images were much deeper than the narrowband images, little noise is added by this subtraction procedure.

To determine the conversion between measured count rates and physical fluxes observed through the narrow-band filters, we have found it useful to use the online Exposure Time Calculators (ETCs)  for WFC3 UVIS and IR\footnote{See {\tt http://www.stsci.edu/hst/wfc3/tools/etcs/;} Derivation of count rates for each object of interest is described in the next section.}.  The results obtained this way were consistent with estimated conversion factors obtained using the PHOTFLAM parameters and filter widths posted in \hst\ on-line documentation, although systematically larger by 5-10\% for the various filters.   We attribute the differences to the fact that the ETCs allow the user to specify the position of the emission line within the filter bandpass, rather than assuming an average throughput over the filter.  We assumed a redshift for M83 of 500 $\VEL$ to position the lines \citep{crost02}.\footnote{Note: the range of velocities observed across the disk of M83 is roughly $\pm100 ~ \VEL$ with respect to this redshift; see \citet{crost02}.}   The conversions derived by this process are shown in Table \ref{table_conversions}.

\begin{figure}
\plotone{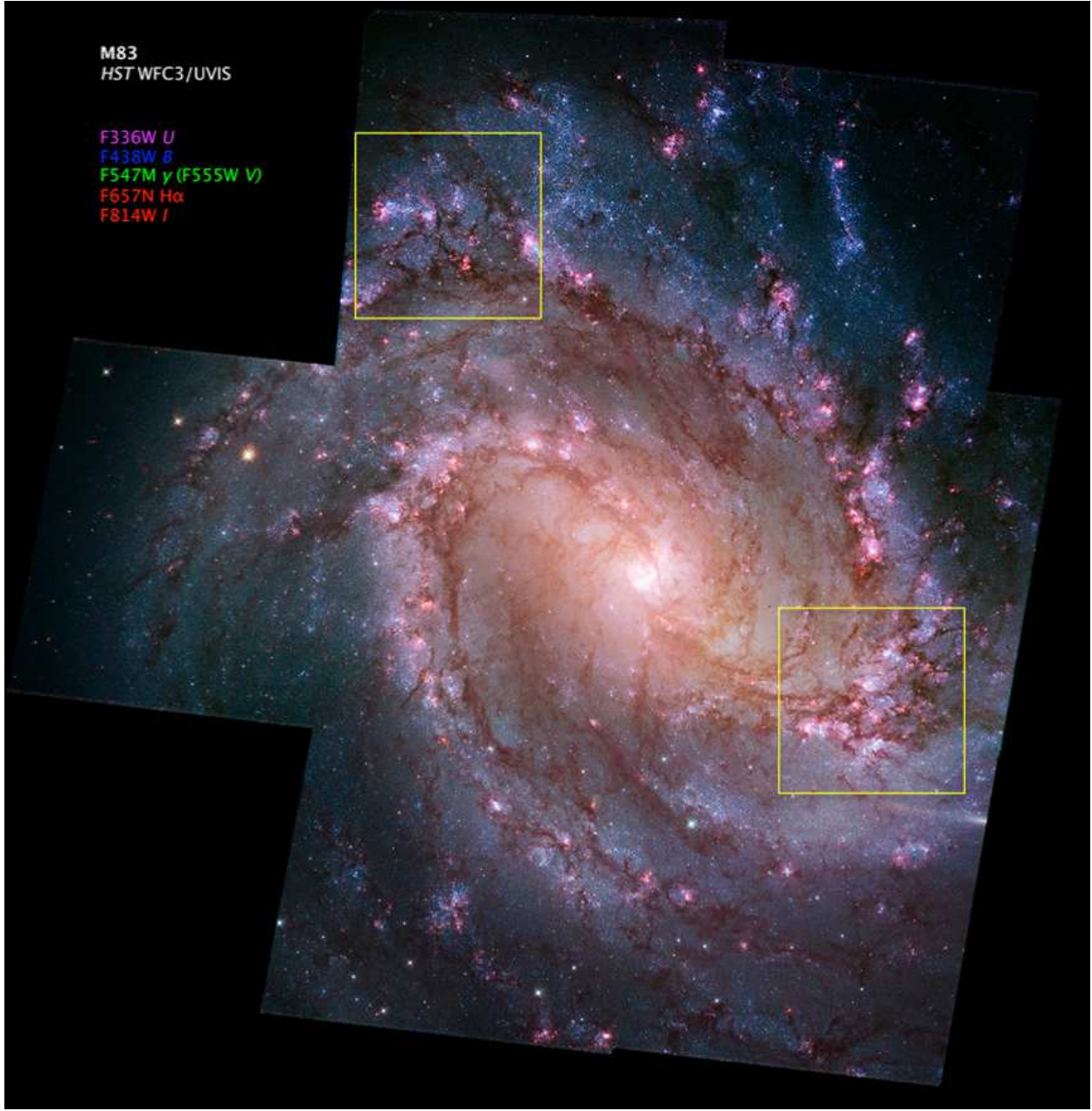}
\caption{A full color mosaic of the seven UVIS fields imaged by WFC3. The legend shows the color coding used for the various filters combined into this mosaic.  The two yellow boxes are 80\arcsec\ square and indicate the regions enlarged in the following two Figures. \label{fig_overview2}}
\end{figure}

\begin{figure}
\plotone{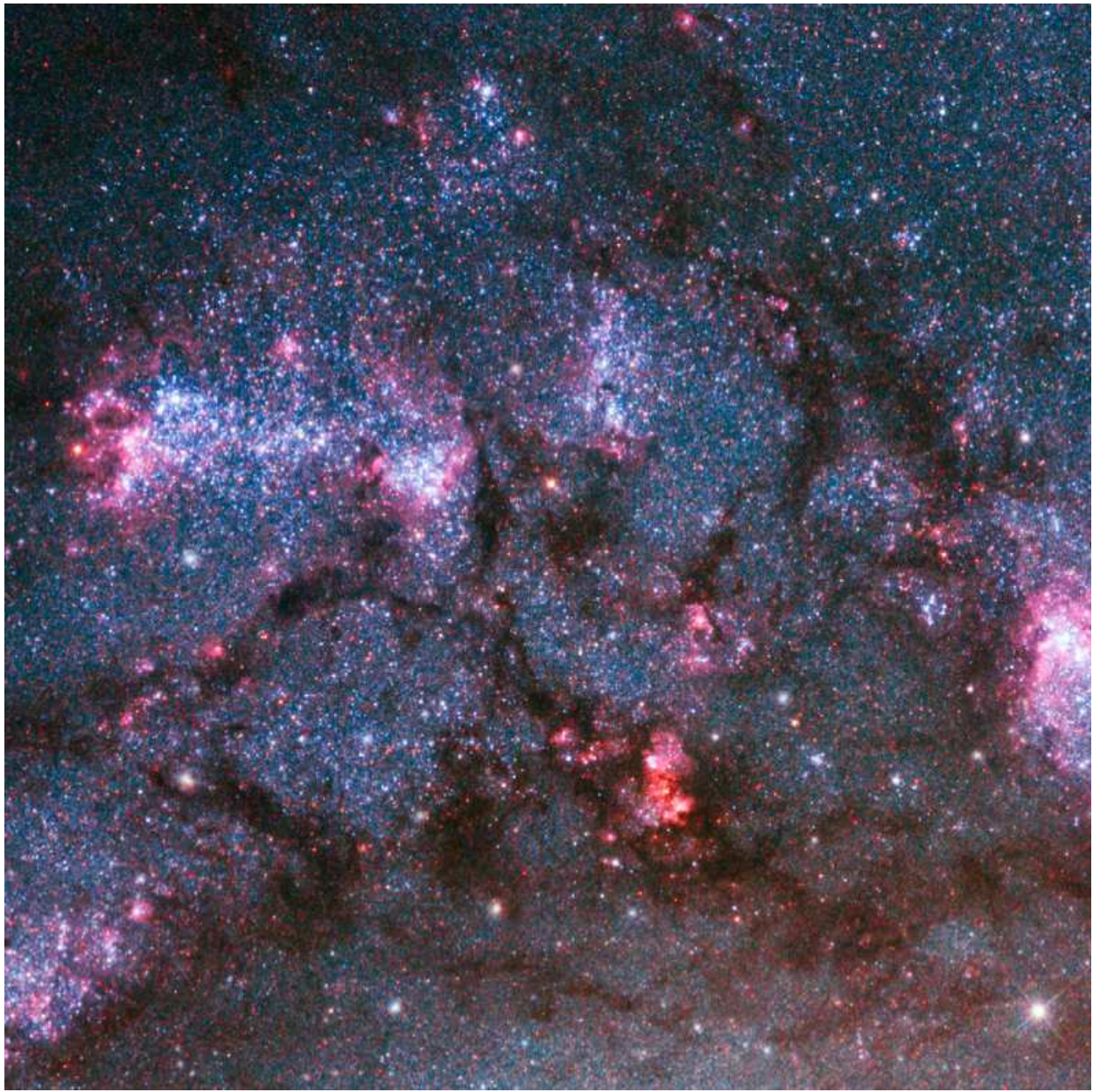}
\caption{A full color mosaic of a portion of Field 2 in the northern spiral arm. The color coding is the same as Fig. \ref{fig_overview2}. The region shown is 80\arcsec\ (1.75 kpc) on a side. The region contains several young star-forming regions, a number of compact star clusters, and a delicate network of dust lanes.  While the dust lanes can be quite opaque, many regions between the dust lanes are nearly free of extinction. \label{fig_nne}}
\end{figure}

\begin{figure}
\plotone{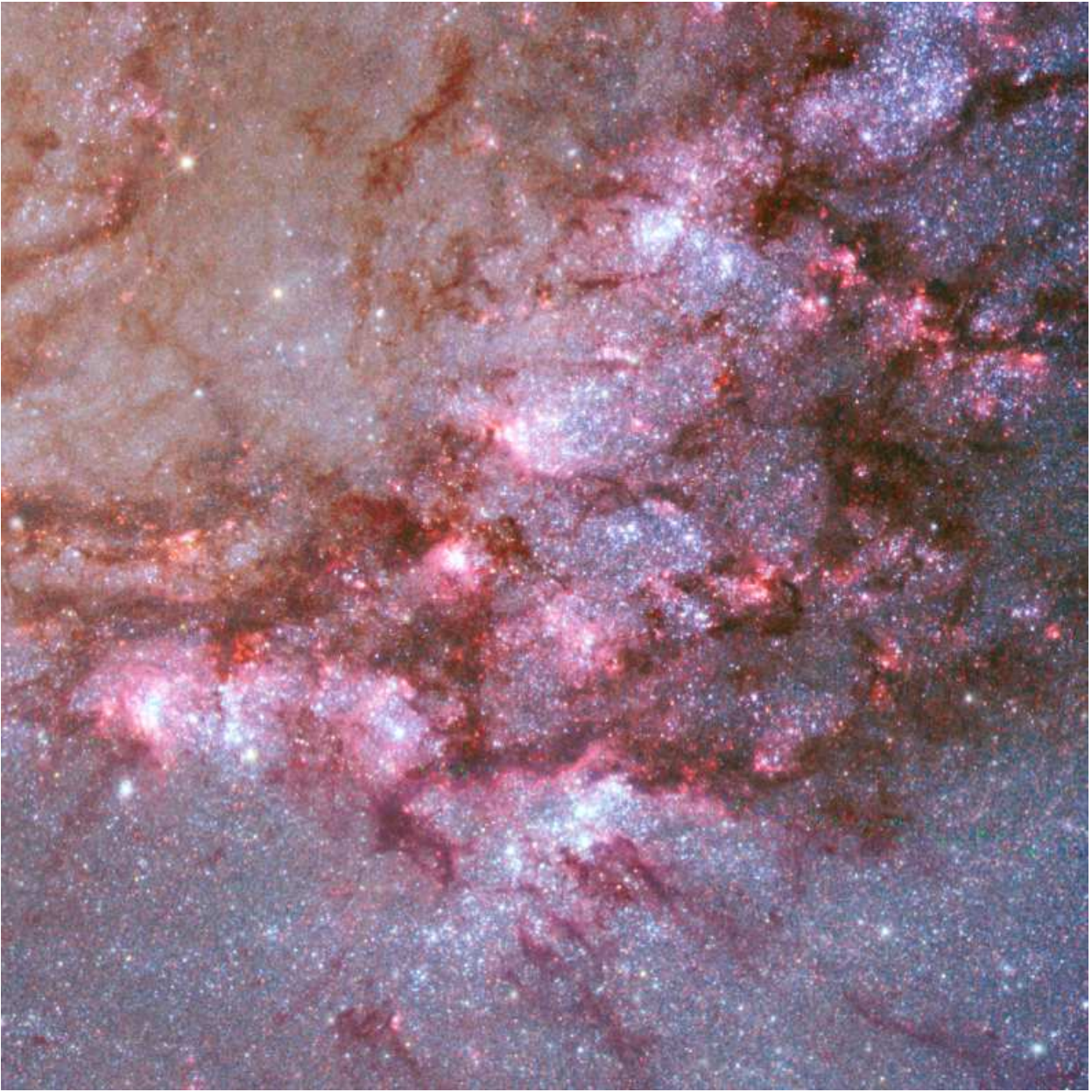}
\caption{A full color mosaic of a huge star-forming mega-complex SW of the nucleus, straddling the border between Field 5 and 6.   The color coding is the same as Fig. \ref{fig_overview2} and the region shown is 80\arcsec\ (1.75 kpc) on a side.  The older bulge population colors the upper left portion of the field shown here, and many regions of young star formation are indicated by \ha-emitting clumps. \label{fig_sw}}
\end{figure}

In Figure \ref{fig_overview2}, we show a mosaic of all seven WFC3 UVIS fields, constructed for us by Zolt Levay (STScI) that provides an overview of the full region covered.  This is an approximately true color image, with U, B, V[y], and I bands showing the stellar component, and the F657N \ha\ also in red, showing the gaseous component.  It is difficult to appreciate the full quality and detail in this overview image without additional magnification.  The yellow boxes in Figure  \ref{fig_overview2} show two 1.3\arcmin\ (1.75 kpc) square regions that are enlarged in Figures \ref{fig_nne} and \ref{fig_sw}, where some sense of the exceptional spatial resolution is more evident.  Further enlargements of smaller fields will be shown in subsequent figures.  Of particular note in these wide-field figures, however, is the appearance of the dust lanes, which show considerable structure, especially in the U and B bands where the attenuation of background starlight is most significant.  However, regions away from the dust lanes are almost clear by comparison.

\section{Identifying SNRs in M83: The \hst\ Advantage \label{sec_find}}

We now focus on an aspect of interest to our team that can be pursued with these data: the characterization of the SNR population in M83.  Below we describe:  a) how we have identified the subset of previously identified SNRs for which the \hst\ data are most advantageous, those being the objects with characteristic sizes below 0\farcs5; and b) how we have identified an extended set of SNR candidates beyond the previous sample based on the \hst\ data themselves and comparison with the \chandra\ data from \citet{long14}.  We choose an angular size limit of 0\farcs5 for objects reported in this paper because objects below this size are extremely difficult to characterize from ground-based measurements but are accessible to \hst.  Table \ref{table_snrlist1} and Table \ref{table_snrlist2} list the basic supporting information for 63 objects selected for the present study using these ground rules.  Details of the information in these tables will be described below.
 
The techniques employed to find SNRs in nearby galaxies has been discussed by  \citet{blair97}, \citet{blair04}, and \citet{long10}, and references therein.  B12 (also refer to the Erratum, \cite{blair13}) discuss a recent ground-based optical search for SNRs in M83.  Since many aspects of the technique used here are essentially identical to what was used by B12, we refer the reader there for details.  

The B12 catalog has  225 SNR candidates based on observed high F(\sii)/F(\ha) ratios in images from the Magellan-I 6.5m IMACS survey, obtained in conditions of $\simeq 0\farcs5$ seeing.  They also list 36 sources found with an alternate technique that has specific application to M83: the F(\oiii)/F(\ha) ratio.  At the super-solar abundances prevailing over much of the  bright optical  disk of M83, \hii\ regions typically have rather low electron temperatures \citep{dopita13}. This means that the collisional excitation rate to the level giving rise to the \oiii\ $\lambda\lambda$ 5007, 4959 lines is very much reduced. Thus, normal \hii\ regions in M83 emit very little \oiii\ over most of the inner disk although this changes systematically in the outermost disk regions.  Compact nebulae emitting substantial \oiii\ emission are either a) normal ISM-shock dominated SNRs with velocities high enough to excite \oiii, b) ejecta-dominated young SNRs akin to Cas A in our Galaxy \citep{fesen01}, c) planetary nebulae (unresolved \oiii\ point sources with no X-ray emission), d) Wolf-Rayet nebulae (slightly extended, with obvious continuum source), or e) an X-ray binary  that can ionize local gas to O$^{++}$.  Because of the SN history of M83, we were particularly interested in finding any examples of young ejecta-dominated SNRs, which have additional diagnostics that include detectable soft X-ray emission and/or high velocities in the spectra of candidate objects.  In principle, B12's \oiii-source list may contain representatives of each of these categories, with the exception of PNe for which we had enough criteria (unresolved \oiii\ sources with no X-ray emission) to remove them from further consideration.

The tremendous advantage provided by the \hst\ WFC3/UVIS data even compared with the excellent Magellan data is in resolving very compact emission regions.  At 4.61 Mpc, the $0\farcs0396$ pixel size of UVIS corresponds to just under 1 pc.  In Field 1, D10 found 60 SNR candidates where only 12 were known in the earlier ground-based study by BL04.  Roughly one-third of these new candidates were in the crowded nuclear starburst region that was smeared out at ground-based resolution.  Compared with the recent SNR catalog of B12, 25 of the 40 non-nuclear SNRs from D10 were found independently, so a less dramatic improvement in the overall number of SNRs found with WFC3 is to be expected compared with B12.  

The additional advantage of the \hst\ data is in its ability to provide quality assessments of sizes and morphologies of the SNRs.  In particular, WFC3 imaging is excellent for finding and characterizing the smallest angular size objects.  However, with the modest exposure times employed here, the larger, lower surface brightness objects tend not to have very many counts per pixel above background.  Fortunately, these larger objects ($>0\farcs75$ or $>$15 pc) are just the ones that the Magellan data are most effective for characterizing. Hence, the Magellan and \hst\ data sets complement each other.

The other important aspect of \hst/WFC3 is in its access to the near-IR.  \citet{oliva89} demonstrated the elevated F(\FeiiL)/F(Br$\gamma~2.166~\mu$m) ratio derived from galactic SNRs compared with \hii\ regions, a point that has been made more recently by \citet{koo13}.  \citet{greenhouse97} used \fe2\ line strengths in M82 in combination with radio data to investigate young SNRs in this dusty starburst.  \citet{alonso03} and \citet{labrie06} have also used \FeiiL\ to investigate local starbursts, finding both point-like and diffuse components of emission, but not tying these to optical shock diagnostics.  Ground-based surveys of nearby galaxies such as M33 \citep{morel02} have detected a few SNRs, but not enough to characterize the \fe2\ emission relative to optical line strengths to understand its full diagnostic power. We have employed the F164N filter to capture the \FeiiL\ emission characteristics of compact nebulae in M83, finding not only a number of the Magellan SNRs from B12, but also new objects not found in that survey.  This is described in more detail below. 

For the current initial effort, we have not performed a complete and systematic search for SNRs in all seven fields.  Rather, we have carried out a targeted visual inspection of the data for each field making extensive use of the SAOImage DS9 display program \citep{joye03} and the following basic technique. We displayed the aligned full field data for each field in two `RGB' frames: in one, we displayed the subtracted \ha\ (R), \sii\ (G), and \oiii\ (B) data and in the other we displayed the I-band (R), V[y]-band (G), and B-band (B) data to show the stellar component.  Another RGB frame was loaded with the \chandra\ data from the recent deep survey by \citet{long14}, with soft X-rays (0.35 -- 1.1 keV) in R, medium (1.1 -- 2.6 kev) in G, and hard (2.6 -- 8 keV) in B.  This allowed us to assess not only whether X-ray emission was present at a given position, but to first order whether the emission was soft and (likely) thermal or bright and/or hard as might be expected from X-ray binaries or background sources.  

Finally, we also displayed the subtracted \fe2\ data as a separate black \& white frame.  With all of the data displayed, we then projected `region' files for all of the known SNRs (B12 and D10) as well as region files showing the positions of the detected X-ray sources from \citet{long14}. We could then zoom in as appropriate and align all of the frames on each source of interest.
This allowed us to carefully inspect the \hst\ data at the positions of soft X-ray sources that had not already been identified as possible SNRs by the Magellan survey, and we could pay particular attention to any compact nebulae that appeared in the \fe2\ data, whether or not they aligned with known sources.  Thus, we effectively have performed a targeted search using the \chandra\ data and the \fe2\ images to confirm and extend the Magellan survey results.
 
We manually adjusted the DS9 regions to a size that was representative for each object.  For most objects that were either close to circular or partial shells, we fit circular regions. For highly asymmetrical objects, elliptical regions were defined.  The J2000 coordinates and radii (in arcseconds) for each object as so defined are listed in Table \ref{table_snrlist1}, along with a conversion to diameters in parsecs at the assumed distance.  (For elliptical regions, we used an average of the major and minor axes to be representative.)  The selection of a diameter of 0\farcs5 as the size cutoff for previously known SNRs in this paper was somewhat arbitrary, corresponding approximately to the seeing obtained in the Magellan ground-based survey of B12.  There were a number of objects that had measured diameters slightly above 0\farcs5 in the \hst\ data that were excluded here, but will be reported in a future exploration of these data.  For the 26 SNR candidates reported here for the first time, a size criterion was not imposed as they are all breaking new ground.  Many of these are below 0\farcs5 in size, but a few of the new objects found in confused regions using the \fe2\ diagnostic are larger than 0\farcs5 in diameter.  We have also calculated and listed the galactocentric distances (GCDs) of each object, based on the same assumptions for galactic inclination and major axis angle as used in B12 and the assumed distance of 4.61 Mpc.

Of the 63 objects reported here, 27 are cross-referenced to objects in the B12 Magellan catalog (five of which are also in D10), 15 overlap with objects from D10 (including the five overlaps with Magellan), and 26 objects are newly discovered.  We note that the D10 objects include some objects from their Table 2 on non-nuclear SNRs, and some from their Table 3 of SNRs in the nuclear region.  We only include objects from the ``nuclear'' list that are actually outside the very bright central core of the galaxy, and that could be seen and measured with little uncertainty.  Hence, the D10 possible Cas A-like object is included here (our object \#38 which is B12-321 and X-ray source X243) and the SN1968L candidate is not. 
We anticipate a separate paper on the multi-wavelength characteristics of the nuclear region itself, including a re-examination of the SNR population in the nucleus.  Cross referencing to SNRs from the earlier lists is provided in column 2 of Table \ref{table_snrlist1}. A column is also included indicating which WFC3 field's data were used for the measurements.  (Some objects appear in the overlap region between multiple fields.)

In order to extract integrated count rates, we defined regions that included all of the visible emission from each object while excluding contaminating emission as much as was possible.   Because many of the objects are embedded in H~II emission of varying brightness, a set of separate background regions for each object were also defined.  Background regions were selected by referring to a display of all four subtracted emission line frames for each field and choosing an appropriate region that worked for all filters.  In general, background regions were larger than the object extraction regions to ensure than the average noise levels were well-sampled.  We wrote a python program to use the regions defined in this way to go into each data set and extract total, background-subtracted net count rates for each object of interest.  The net count rates were then used with the conversion factors shown in Table \ref{table_conversions} to derive observed, integrated, background-subtracted fluxes in each filter for each object.

Table \ref{table_snrlist1} lists the derived continuum-subtracted F657N fluxes,\footnote{Note: several objects found via \fe2\ emission but with no significant optical emission list the F164N flux in the notes column instead.} which includes both \ha\ and \nii\ lines, and shows the calculated ratio of F673N flux to the F657N flux, which provides a lower limit to the ratio of F(\sii)/F(\ha).  Because of the unknown impacts of \nii\ on a case-by-case basis, it is not possible to set a hard limit on this ratio of filter fluxes that may correspond to F(\sii)/F(\ha) = 0.4, the usual criterion for indicating shock heating that dates back to \citet{mc72} and \citet{sabba76}.  Existing SNR spectra in M83 (BL04) indicate that \nii\ is very strong and that only roughly 1/3 of the F657N flux is due to \ha\ (somewhat dependent on GCD).  Additional spectra will be needed to quantify this further for our sample.
In any event, the selected objects all have elevated ratios compared with local \hii\ region emission in these data.  Also, as shown by the comments in Table \ref{table_snrlist1}, essentially all of the objects registering lower ratios show significant \hii\ contamination upon visual inspection of the data.

In Table \ref{table_snrlist2}, we expand the reporting of derived flux information, repeating the F657N flux for reference, but then going on to report various other filter ratios.  We also indicate which objects have X-ray counterparts 
as given by \citet{long14}. Some 32 objects have likely X-ray counterparts, although the positional agreement for a few (shown with parentheses around the X-ray source ID) are more uncertain and two objects (\#23 and \#24) are so closely spaced that it is not clear which (or whether both) correspond to the X-ray source.  More detailed comments based on a visual inspection of each source are provided in the last column of this Table.  

Scanning column 5 of Table \ref{table_snrlist2}, the F(F164N)/F(F657N) ratio varies by more than a factor of 100.  Because of the uncertain but substantial and variable effect of \nii\ contamination in F657N from object to object, this ratio is a poor representation of the actual F(\fe2)/F(\ha) ratio.  Assuming \ha\ is responsible for 1/3 of the F657N flux (cf. BL04), the F(\fe2)/F(\ha) ratio likely varies from $\leq$0.01 to $\geq$5 in the most extreme cases where no F657N flux from the object is clearly detected.  In column 6 of Table \ref{table_snrlist2}, we also show F(F164N)/F(F673N), i.e. F(\fe2)/F(\sii), and much the same effect is seen but shifted to higher ratios (since \sii\ is generally weaker than \ha+\nii).  The reason for such substantial variations in relative \fe2\ emission, at least at the high ratio end, is likely driven by high levels of extinction, which decreases the optical lines in relation to the IR \fe2\ feature, sometimes to the extent of making the optical counterpart undetected.  
It is less obvious why, given the propensity for readily detectable \fe2\ in so many of the shock-heated objects, there are some that show little or no \fe2.  In the absence of reddening, it is encouraging that the predicted ratios from  {\it MappingsIII} models \citep{allen08} are consistent with the lower end range of our actual observed ratios, although the abundances and other assumptions will need to be adjusted in future calculations in order to be appropriate for the higher abundances and ISM densities in M83.  Spectra are also needed in order to correct the observations for extinction.

Table \ref{table_snrlist2} also lists the F(F502N)/F(F657N) ratio in column 7.  As discussed above, multiplying by 3 to correct for \nii\ contamination provides an approximation to the observed F(\oiii)/F(\ha) ratios, and extinction correction would also increase this ratio even further.  Even so, while many objects have detectable and even substantial \oiii\ emission, only one, \#46 which is the previously known SN1957D, has an extremely high ratio expected for the ejecta-dominated case. 
Even the compact object similar to Cas A that was first identified by D10 on the eastern edge of the bright nuclear region (our object \#38) only has an observed F(F502N)/F(657N) ratio near unity, and in this case, there is \ha\ directly associated with the object (see Figure 4 of D10).  The generally low values we see for this ratio are clearly counter to our expectation that we would find numerous Cas A analogs in this survey of the smaller (hence, younger) population of SNRs in M83.  Possible reasons for this are discussed further in the next section.

Based on the above assessments, we have high confidence that each of the objects listed in Tables \ref{table_snrlist1} and \ref{table_snrlist2} represents a SNR despite the somewhat disparate criteria involved for some of the individual objects.

\section{Discussion \label{sec_discussion}}

The \hst\ data resolve many complex regions of star formation, dust lanes, and nebular emission that are unresolved even at the best ground-based resolutions available.  This resolution provides several distinct advantages for studying SNRs.  Accurate sizes and morphologies can be determined for many of the SNRs first identified in the ground-based data. Objects in complex regions can be more readily identified and separated from contaminating stellar or nebular contamination.  Also, the direct determination of the sizes of the smallest SNRs ($\leq$10 pc in diameter) allows identification of the population of the youngest SNRs.  Many of these SNRs were missed in ground-based surveys because they were either too faint or simply smeared out by atmospheric seeing.  Below we show a few example objects to demonstrate these points, discuss our initial assessment of the small SNR population, and highlight the use of \fe2\ emission as a diagnostic of shocks and SNRs in particular.

\subsection{Selected SNR Candidate Identifications\label{sec_examples}}

\begin{figure}
\epsscale{0.75}
\plotone{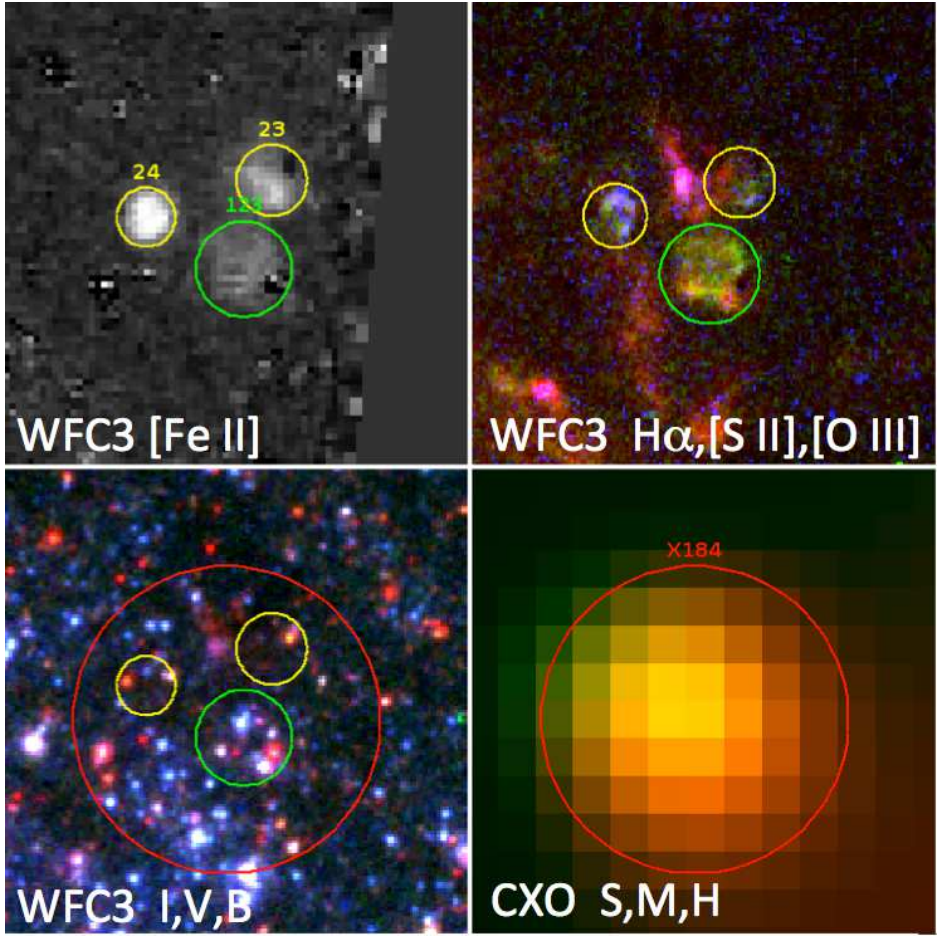}
\caption{This 4-panel figure shows a 6\arcsec\ square region from the Field 2/4 overlap region as an example of where the full power and resolution of \hst\ can be seen. In this and the following two figures, the panels include the subtracted NIR \FeiiL\ emission at upper left, a color optical continuum-subtracted emission line image (R=\ha, G=\sii, B=\oiii) at upper right, a color continuum image (R=I, G=y, B=B) at lower left, and a color version of the \chandra\ data (R=0.35 Ð-1.1 keV, G=1.1 Ð- 2.6 keV,  B=2.6 -- 8 keV) at lower right.  The three circled objects stand out as greenish-yellow and bluish in the upper right panel relative to nearby \hii\  region (red) emission because of relatively strong \sii\ and/or \oiii\ emission, and are thus SNRs.  The yellow circled objects are \#23 and \#24 from our Table 3, and are new from this study.  The green circled object was identified in the Magellan SNR survey (object B12-123), but is larger than 0\farcs5 and so is not listed in the Tables in this paper.    \hst\ resolves the objects, and all are also seen clearly in \fe2, although the relative intensities vary. The X-ray panel shows a somewhat extended X-ray source (X184) covering all three objects, so all could be contributing.  The red circle is 4\arcsec\ in diameter.   All regions are shown on the continuum panel for context.  A number of young blue stars and red supergiants are in the vicinity, but the SNRs likely dominate the X-ray emission. \label{fig_triplesnr}}
\end{figure}

 A small (6\arcsec\ square) region from Field 4 is shown in Figure \ref{fig_triplesnr}, where the four panels show the subtracted \fe2, 3-color continuum subtracted emission lines, 3-color continuum (all from WFC3) and a 3-color representation of the \chandra\ data.  (Details are given in the figure caption.)   Three patches of bright \fe2\ emission are evident, and correspond to three regions of enhanced \sii\ and to a lesser extent \oiii\ emission (causing the yellow-green and/or bluish-white appearance in the upper right panel).  Only the brighter and larger object below center in the figure was identified in the Magellan survey.  The morphology seen with \hst\ resolution and the \fe2\ emission both indicate these are three separate but closely spaced SNRs, two of which were previously unknown.  
 Although a \chandra\ source was identified at this position, it is extended compared to most point sources and may involve emission from more than one of the objects.  The continuum data (lower left panel) show a combination of hot, young stars and red supergiants in the vicinity, indicating the progenitor of the SN arose from within a young population.

\begin{figure}
\epsscale{0.9}
\plotone{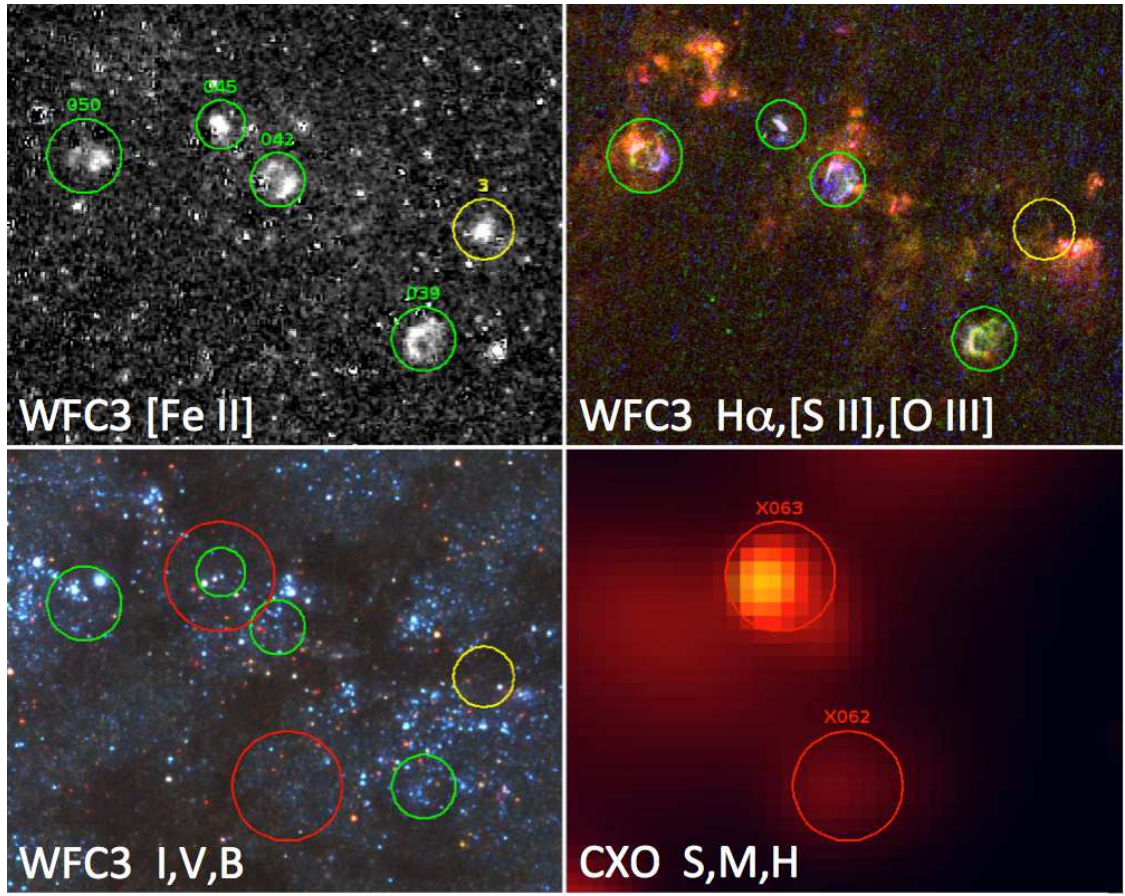}
\caption{This 4-panel figure shows a 16\arcsec\ $\times$ 20\arcsec\ region from Field 5 but the panels and colors are the same as in Figure \ref{fig_triplesnr}.  Green circles indicate four optical SNRs identified in B12 (from left to right, objects B12-50, B12-45, B12-42 and B12-39).  All four of these SNRs are larger than 0\farcs5 and hence do not appear in our tables, but they stand out from nearby \hii\  region (red) emission in panel 2 because of strong \sii\ and/or \oiii\ emission.  All are also seen in \fe2\ emission, although only B12-45 has a clearly detected soft X-ray counterpart.  The yellow circle indicates a new source seen in \fe2\ that is not seen in optical or X-ray.  This is likely a SNR that was missed because of significant foreground dust absorption, as seen in the lower left panel.  The other objects, while close to dust lanes, appear to be in relatively clear lines of sight.  \label{fig_5snrs}}
\end{figure}

In Figure \ref{fig_5snrs}, we show a 16\arcsec\ $\times$ 20\arcsec\ region from the southern portion of Field 5 as another example.  The green circles denote four B12 catalog objects in the field.  
The subtracted WFC3 emission-line data resolves each object into knots, shells, or partial shells, and reasonable measurements of the size of each object can be made.  Although the IR camera resolution is lower, all four of these objects are nonetheless well detected in \fe2\ and show reasonably similar morphologies to the UVIS emission line data.  While stars are present near each object, the stellar contamination is not severe.  The smallest SNR at upper left, B12-45, is well-detected in X-rays, and two neighboring SNRs are within more diffuse X-ray emission but were not identified as separate point sources in the \chandra\ data.  Of particular interest is the yellow-circled object in the \fe2\ panel of the Figure.  This object is comparable in size and \fe2\ surface brightness to the other SNRs, and yet no optical or X-ray counterparts are evident at the position.  The yellow circle projects onto a dark dust lane in the continuum image panel.  We posit that this is a previously unknown SNR, listed as object \#3 in our Tables  \ref{table_snrlist1} and \ref{table_snrlist2}, whose optical (and possibly soft X-ray) emissions are blocked from our vantage point.

\begin{figure}
\epsscale{0.85}
\plotone{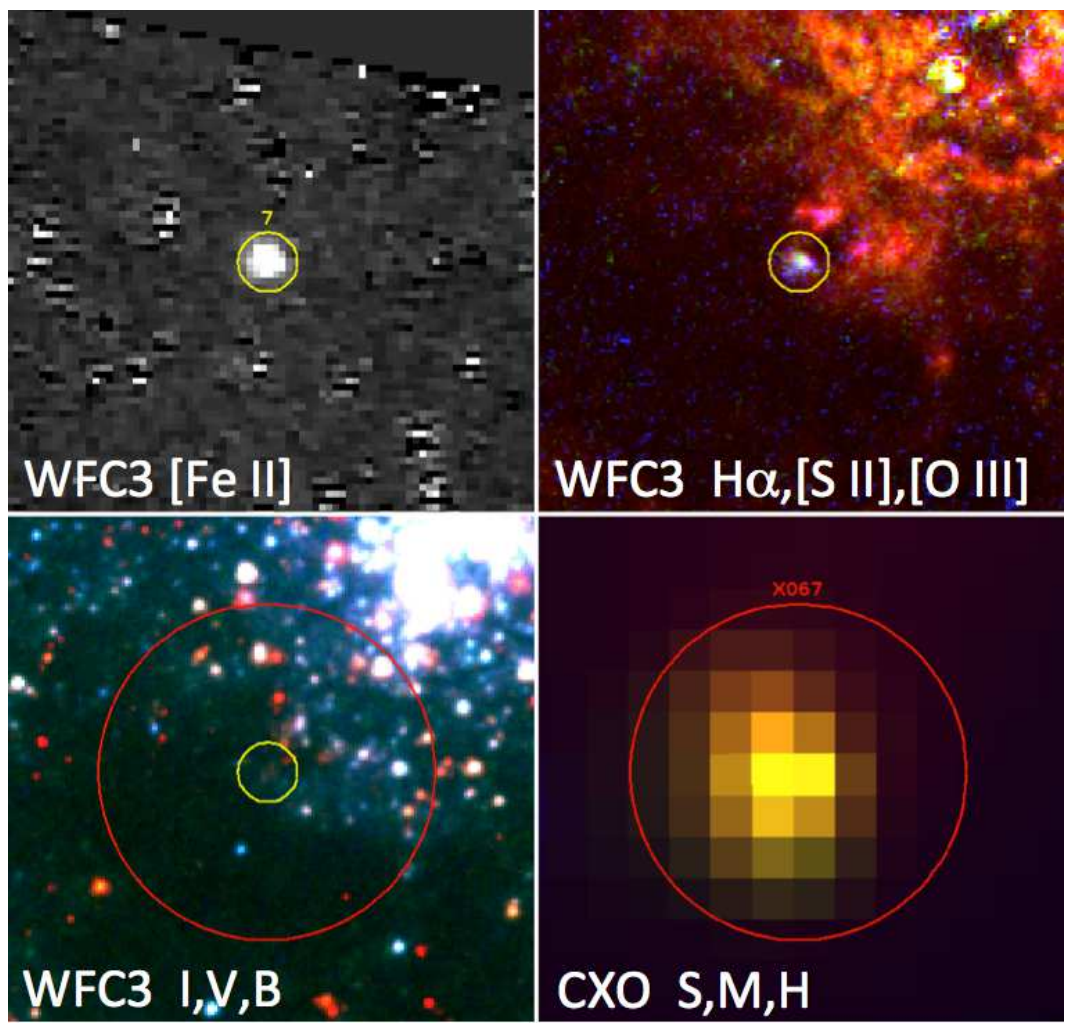}
\caption{This Figure shows a 6\arcsec\ region from Field 5 centered on a previously unidentified very compact young SNR in M83 (\#7 in our list).  The red circle showing the X-ray source ID is 4\arcsec\ in diameter, and the panels are the same as in Figure \ref{fig_triplesnr} above.  The optical remnant is just barely resolved, although hints of more extended \oiii\ (blue) emission may be present. This object has much stronger soft X-ray and \fe2\ emission compared with the faint optical counterpart and appears to be on the edge of the dusty shell surrounding a region of very active star formation visible at upper right in the continuum panel.  The bright \hii\ region (red in panel 2) is completely absent in \fe2, while the SNR emission is very well detected.  
The proximity to recent star formation implies this is likely a young SNR from a core-collapse SN, but the white color in panel 2 indicates strong \sii\ and \oiii\ emission in conjunction with \ha, and thus a radiative shock rather than ejecta.  \label{fig_newsnr}}
\end{figure}

In Figure \ref{fig_newsnr}, we show an example of a source (\#7 in our list)  identified both from its X-ray and \fe2\ emission, located on the northern edge of Field 5.  We were initially drawn to this location by the moderately strong soft X-ray emission (source X067 from \citet{long14}), for which no optical candidate had been identified.  In the subtracted emission-line panel, a very compact knot of optical emission is present with characteristics consistent with a SNR identification, directly adjacent to a bright \hii\ region and active star-forming region. The \fe2\ data show that  the same object is clearly detected while the optically much brighter \hii\ region to the NW is not seen in \fe2.  Because of the larger pixel size for the IR camera, it is not immediately obvious whether only a portion object \#7 is bright optically, or whether the optical emission represents the SNR's true size.  The object appears projected against a fairly dark, dusty region on the outskirts of the star-forming region to the NW, but in this case it must be primarily on the near side of the obscuration, and thus not as heavily reddened as the new SNR highlighted in Figure \ref{fig_5snrs}.  The proximity of this SNR to the star forming region is suggestive of a core-collapse progenitor.

\subsection{Where are the Ejecta-dominated SNRs? \label{sec_noejecta}}

Pertaining to the smallest diameter SNRs, one thing is immediately obvious from Table \ref{table_snrlist2}: while many of the small objects are moderately strong \oiii-emitters, none of them is dominated by \oiii\ emission the way the prototype ejecta-dominated remnants are.  Cas A in our Galaxy (age $\simeq$ 340 years, diameter=5 pc) shows primarily lines of O and S ejecta up to velocities in excess of 8000 $\VEL$ \citep{{kirshner77},{fesen06}} and has no comparable associated \ha\ emission. The Small Magellanic Cloud object 1E0102-7219 (hereafter E0102; kinematic age $\simeq$ 2000 years, diameter=11 pc) is also dominated by high velocity  ($\sim$6000 $\rm km ~ s^{-1}$) O emission lines, and while it is adjacent to an \hii\ region, it has no directly associated \ha\ component \citep{{dopita84}, {blair00}, {vogt10}}.  Thus, to first order, E0102 simply looks like an older (and larger) version of Cas A.  The object N132D in the Large Magellanic Cloud is more than 3000 years old and still shows evidence of O-rich ejecta \citep{vogt11}, although an outer shell swept up by the main blast wave is starting to show regions of radiative shock emission \citep{{morse96},{blair00}}.  However, with a diameter of 25 pc, this outer shell would be readily resolved with WFC3 for a similar object in M83.  \hst\ images of the extraordinary young SNR in NGC~4449 \citep{milis12} seem to show \ha\ as well as very bright \oiii\ emission at the source position, although spectra show the \ha\ to be narrow while the \oiii\ lines are broad and hence, are due to the expanding ejecta.  

\begin{figure}
\centering
\includegraphics[width=0.5\textwidth, angle =90 ]{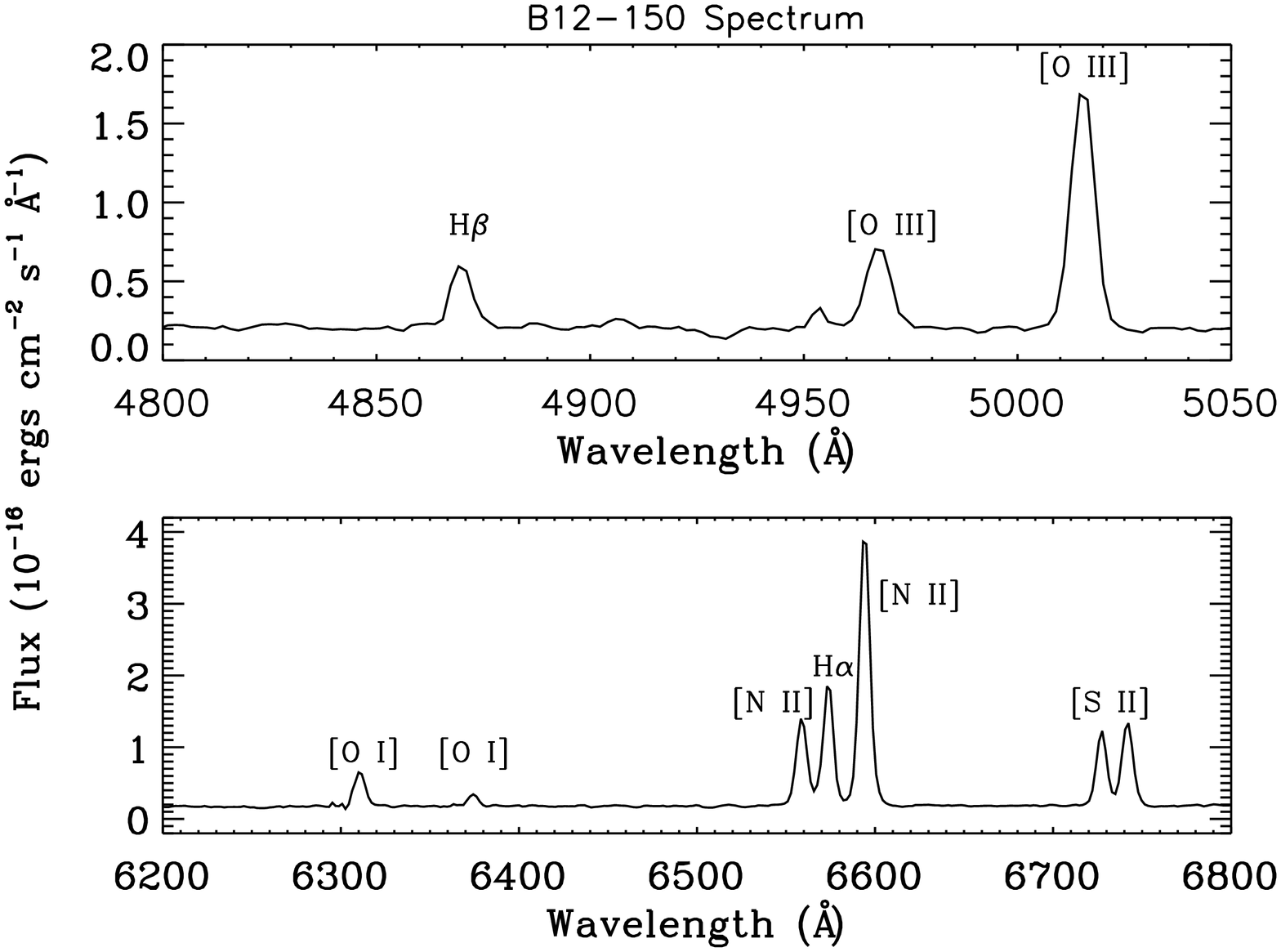}
\includegraphics[width=0.5\textwidth, angle =90 ]{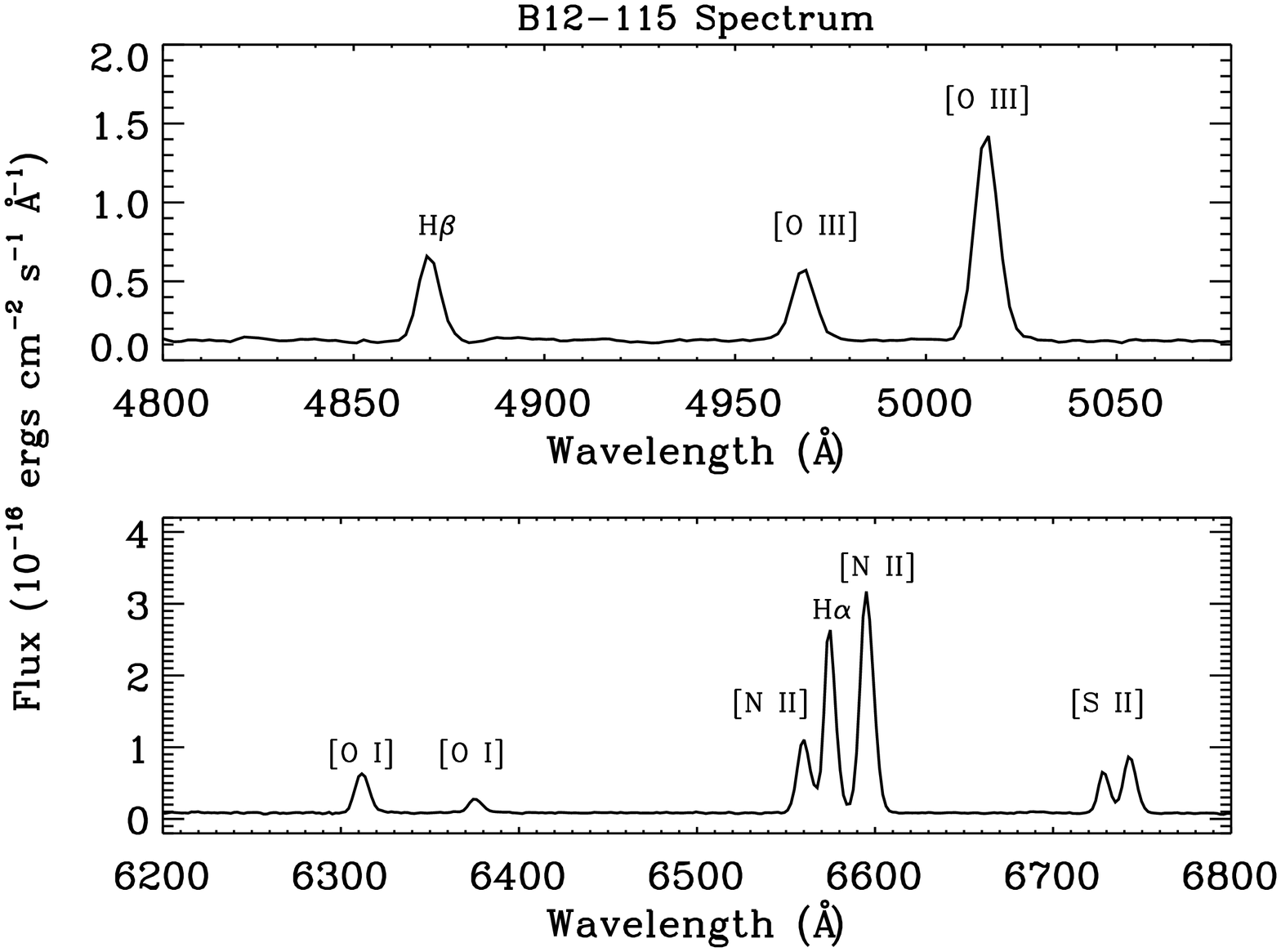}
\caption{This Figure shows Gemini-GMOS spectra of two objects in our small diameter SNR list: (top) B12-150 (\#41), and (bottom) B12-115 (\#18).    Note: the vertical scale changes by a factor of two between the two panels of each spectrum.  Strong forbidden lines and lack of high velocity emission, especially on the \oiii\ lines, points toward radiative ISM shocks rather than ejecta emission.
\label{fig_spectrum}}
\end{figure}

The fact that in our M83 imagery both \ha\ and \sii\ are comparable to the \oiii\ emission in nearly all of the small diameter SNRs points toward emission from normal radiative ISM shocks rather than an ejecta-dominated interpretation, which is not what was expected for these small diameter (presumably young) SNRs.  Spectra are currently only available for selected objects from our list in Table \ref{table_snrlist1}, but they appear to corroborate the conclusion from imagery.  In Figure \ref{fig_spectrum}, we show spectra of two objects obtained as part of an ongoing Gemini-S GMOS program on M83 (P. F. Winkler, PI) that will be reported separately.  B12-150 (\#41 in our Tables), which was also observed spectroscopically by BL04 (coincidentally also their object BL04-41) has a diameter of 11.3 pc, similar to E0102. However, the spectrum of this object is fully consistent with ISM-dominated radiative shocks: \oiii\ $\lambda$5007 is comparable in strength to \ha\ with no evidence of high velocities that would indicate emission from ejecta.  This object is only 0.8 kpc from the center of the galaxy and has very strong \nii\ and \sii\ lines compared with \ha.  The \sii\ doublet line ratio of 0.85 implies high electron densities near 700 $\rm cm^{-3}$.

The second spectrum in Fig. \ref{fig_spectrum},  B12-115 (\#18), poses an even more extreme example.  Only 5.3 pc in diameter, it is essentially the same size as Cas A, and yet its spectrum shows no signs of broad ejecta emission lines.  This object is 1.8 kpc from the center of the galaxy, and shows somewhat weaker lines of \nii\ and \sii\ compared with \ha, although some modest contamination by nearby \hii\ emission may be present in the spectrum.  As with B12-150, this object has  moderately strong but narrow \oiii\ emission consistent with bright radiative shocks.  The \sii\ doublet line ratio of 0.75 implies an even higher electron density of about 1000 $\rm cm^{-3}$ in this case.

Published spectra for a number of M83 SNRs were provided by BL04, but only two are from the current ``small SNR'' sample, the aforementioned B12-150, and B12-147 (our \#40, also BL04-40).  The BL04 spectrum of B12-150 is consistent with the higher quality GMOS spectrum discussed above.  The B12-147 spectrum is also consistent with a classic ISM shock, with strong forbidden line emission no evidence of high velocity ejecta.  An interesting point is that, once again, the electron density for B12-147 implied by the \SiiL\ ratio of 0.97 is moderately high at 300 $\rm cm^{-3}$.  

From the evidence at hand, we conclude that the conditions in M83 are such that the young SNR population is evolving very quickly beyond the ejecta-dominated phase and into the radiative phase.  A detailed assessment of this finding is beyond the scope of this initial report, but it seems likely that the high pressure/high density ISM conditions and quite possibly the high elemental  abundances in M83 are both contributing to this situation.  In a statistical analysis of Field 1 SNRs and comparison to shock models, D10 found indications of high pressure ISM conditions, both generally and especially within the spiral arms.  The bright, diffuse X-ray emission seen with \chandra\ \citep{long14} is another indication of this high pressure ISM, as is the \ha\ luminosity function for M83 SNRs reported by B12 and shown corrected in \cite{blair13}, which is offset toward significantly higher luminosities than seen for SNRs in M33.  The high \sii\ electron densities in the spectra reported above are further evidence that the shocks in these young SNRs are encountering relatively dense surrounding interstellar or circumstellar material.  The elevated abundances in M83 may contribute by enabling enhanced stellar wind mass loss from the precursor stars \citep{{vink01}, {kudrit02}}, and it could be this material, at least in some cases, that is responsible for the dense, radiative shocks inferred for these young SNRs.  

SN1957D (our object \#46) remains the only confirmed case of a young ejecta-dominated remnant in M83, but it may provide a clue to young SNR evolution in M83.  This $\sim$56 year old SNR has already experienced substantial decline of its ejecta-dominated optical emission over the last two decades \citep{{milis12}, {long12}}, very different from Cas A, which is still bright (and even brightening) more than 300 years after the explosion.  In Cas A, dense ejecta knots light up optically as they encounter the reverse shock, and then fade.  The overall brightening indicates more new ejecta knots are encountering the reverse shock than are disappearing as they cool below detectability. The rapid fading of the broad O lines for SN1957D is consistent with the idea that the bulk of the ejecta have already encountered the reverse shock and that overall the ejecta emission is decreasing.  This indicates that the ejecta-dominated phase will be short-lived compared with objects such as Cas A and E0102, which is consistent with the picture outlined above.

\subsection{[Fe II] as a Shock Indicator \label{sec_iron}}

The WFC3/IR data using the F164N filter provide an excellent tool for locating shock-heated gas, and many SNRs in particular.  As we inspected the seven fields of \hst\ data, the only compact optical nebulae that aligned with \fe2-emitting sources were D10 or B12 SNRs.  Many (but not all) of the known SNRs, including both small diameter SNRs and larger objects not reported here, were found to be strong \fe2\ sources.  One exceptional object in the \hst\ data just to the NE of the bright nucleus was initially identified by D10 as a nuclear SNR candidate (object 16 in Table 3 of D10).  However, a new assessment \citep{soria14} indicates this object is a microquasar similar to SS 433/W50 in our Galaxy; it is a strong X-ray and ATCA radio source and it is exceptionally bright in \fe2.  Hence, even though it may not be a SNR in the usual sense, this object clearly involves a strong shock-heated emission component, consistent with our findings for the more conventional SNRs.  

In support of this conclusion, we examined the positions of dozens of PNe, both from the listings of \citet{herrmann08} and  \citet{herrmann09}, and from our own inspection of objects with similar character in the \hst\ data but not in those listings, and found no \fe2\ emission from any of them.  We also compared to many dozens of W-R stars/nebulae cataloged by \citet{hadfield05} (see their Appendix A, Tables A1 and A2), and found no \fe2\ counterparts.  Occasionally one can see very faint, diffuse \fe2\ emission associated with the positions of very high surface brightness \hii\ regions, but \hst\ is not the right tool for assessing faint diffuse emission.  If M83 were more distant and unresolved, this faint diffuse emission could conceivably compete with or even dominate a global assessment of \fe2\ emission.  \citet{alonso03} and \citet{labrie06} concluded that point \fe2\ sources (e.g. SNRs) accounted for up to a few 10's of percent of the total \fe2\ emission in a number of nearby starburst galaxies, and that the overall \fe2 emission was a good indicator of the star-formation rate and supernova activity.  It would require deeper integrations with the F164N filter in M83 to obtain a quality measurement of the total \fe2\ emission.

As we pursued our targeted search of X-ray source positions, we occasionally found corresponding compact \fe2\ emission sources that were not previous optical SNR identifications.  On closer inspection of the \hst\ optical emission line data at these positions, faint new optical counterparts consistent with an SNR identification were sometimes found, as was the case with the object shown in Figure \ref{fig_newsnr}.  Also, some new compact  \fe2\ sources were found buried in bright or confused regions of \ha\ emission; these are likely buried SNRs for which the usual optical diagnostics for shocks are less effective.  Finally, a small number of well-detected \fe2\ sources did not have obvious optical or X-ray counterparts but were projected against regions of high extincttion), as with the SNR highlighted by the yellow circle in Figure \ref{fig_5snrs}.  Hence, \fe2\ provides an effective way of finding SNRs that are difficult to locate using the usual optical and X-ray diagnostics, thus permitting a more complete sample to be assembled.

As discussed in section \ref{sec_find} and shown in Table \ref{table_snrlist2}, significant variations in the ratio of \fe2\ to optical lines is observed, but until spectroscopy and/or other means of estimating extinction to individual objects becomes available the intrinsic variations in the strength of \fe2\ relative to the optical lines cannot be assessed.  This is because of the differential extinction between optical and IR wavelengths that has the effect of artificially enhancing the observed relative \fe2\ line strength. For instance, the spectrum of B12-150 in Figure \ref{fig_spectrum} shows an observed ratio of F(\ha)/F(\hb) = 5.0 which, using a standard extinction curve \citep{cardelli89} with R=3.1, implies an extinction of $A_V$= 1.64.  Using \sii\ as a clean reference (since \ha\ is not measured directly by F657N), the observed ratio F(\fe2)/F(\sii) = 0.29 corrects to an intrinsic ratio of 0.11.  For other objects with more or less extinction than B12-150, the correction would obviously be larger or smaller.  Hence, quantitative comparison to shock models cannot yet be done for most of our objects.\footnote{The other object with a spectrum in Figure 8, B12-115, coincidentally has nearly the same extinction as B12-150, but B12-115 was not covered in the WFC3/IR observations.} 

In contrast with the reddened objects that have artificially high observed \fe2\ emission are some optical SNRs that have little or no \fe2\ emission detected.  Columns 5 and 6 of Table \ref{table_snrlist2} show some objects with only upper limits on their \fe2\ emission even though one or more optical lines are well detected.  In addition, there are some fairly bright optical SNRs in the sample of larger diameter SNRs not reported here that have little or no detectable \fe2\ emission. Hence, we can say that variable extinction is not the only cause of the observed variations in the relative strength of \fe2\ to the other lines.  

The observational result that there are optical SNRs with little or no detectable \fe2\ indicates there is some region of parameter space that produces relatively weak \fe2\ emission compared to the optical lines.  There are many factors that could contribute to the relative strength of \fe2\ emission, including variable shock conditions, variable amounts of dust destruction in the SNR shocks, the possible contribution of Fe ejecta emission (or not), in addition to variable extinction.  We will investigate these possibilities further once the larger SNR data set has been analyzed more fully, including additional spectroscopy of many of the objects to properly account for extinction effects.  For now we conclude that \fe2\ emission provides an important new diagnostic for identifying shock-heated nebulae, especially in dusty or confused regions, but that it is not a universal diagnostic.

\section{Summary \label{sec_summary}}

We have performed a detailed imaging survey of seven \hst-WFC3 fields covering much of the bright optical disk of M83 in nine continuum and emission line bands. We describe the acquisition and processing for the entire data set and then discuss results of a preliminary targeted search of these data for supernova remnants.  Comparisons between deep \chandra\ X-ray images and the \hst\ data have directed us to a number of new SNRs missed in ground-based surveys but that are apparent at \hst\ resolution.  As part of this search, we have also found the WFC3/IR data using the F164N filter of particular interest since the \fe2\ emission from many SNRs makes them stand out clearly even in dusty or optically-confused regions of the galaxy.  We have also inspected the positions of SNRs previously known from ground-based surveys and selected those that are measured with \hst\ to be less than 0\farcs5 (11 pc) in diameter, and thus must represent the younger population of SNRs in M83. 
In total, we tabulate information for 63 SNRs, 32 of which have an associated X-ray source in the deep \chandra\ survey data \citep{long14}.  

The observed line ratios for this young SNR population, mostly from the \hst\ imagery but with some spectral confirmation, indicate the objects are dominated by radiative ISM shocks rather than ejecta, which was not expected {\it a priori}.  This may be related to both the super-solar abundances, which allow more significant winds and mass loss from the precursor stars, and the apparently high-pressure ISM in M83 which helps to constrain this material to the vicinity of the SN.  The young SNR shocks then encounter this dense material, causing the apparent rapid evolution into the radiative stage.  This interaction would also create a strong reverse shock and the optically-emitting ejecta must pass through this shock and fade quickly in comparison to local examples of young SNRs such as Cas A and E0102.  Additional spectra of this small SNR population, and in particular some of the objects newly discovered with \hst, might be able to find additional transitional cases similar to SN1957D, where the rapidly-moving ejecta are still visible.  This would help confirm this picture of rapid evolution for these young SNRs.

When a more thorough and systematic search is completed for the new fields, and when the complex nuclear region is included (an additional 19 SNR candidates already known; see D10), we expect the total SNR population in M83 to top the 300 mark.  One of several future papers will provide a more complete analysis of the entire SNR population as seen by \hst, including an analysis of the stellar populations near many SNRs to constrain the main sequence turn-off masses of associated stars and hence constrain the masses of many of the SN precursors.

\acknowledgements

We thank Zolt Levay (STScI) for producing the mosaicked image in Figure 2, and Sylvia Baggett and Beth Perriello (STScI) for excellent technical assistance in preparing the \hst\ program for execution during the period when CTE mitigation strategies were just being devised.  
We also thank our colleagues on the companion multi-wavelength surveys of M83 (\chandra, ATCA, and Jansky VLA), including Paul Plucinsky, Chris Stockdale, Leith Godfrey, James Miller-Jones, and Aquib Moin, for useful discussions and ongoing support.
WPB acknowledges STScI grant GO-12513-01 to Johns Hopkins University, and support from the Dean of the Krieger School of Arts and Sciences and the Center for Astrophysical Sciences at Johns Hopkins University during this work.

\bibliographystyle{apj}

\expandafter\ifx\csname natexlab\endcsname\relax\def\natexlab#1{#1}\fi

\section{Appendix \label{sec_appendix}}

Section \ref{sec_obs} provides an overview of the \hst\ data and processing that are used in this paper.  This Appendix provides an expanded description of the data processing steps used to produce the aligned \hst/WFC3 data.  
Specific information on the operation of DrizzlePac software suite can be found in \citet{Fruchter10} and \citet{Gonzaga12}.
Because of the extensive effort required to systematically process this entire data set to a common standard, and because these data have utility for many other science questions beyond those proposed by our team, we are working with the Mikulski Archive for Space Telescopes (MAST) staff at STScI to provide these data to the community as a High Level Science Product.  The availability and location of this resource will be announced directly by MAST.

The {\it tweakreg} task was used to
produce accurately aligned versions of the pipeline-produced FLT images for each 
filter.  The input parameters for {\it tweakreg} were changed as needed for each filter
and/or field in order to optimize alignment. The best configuration was
chosen from the fits resulting in the lowest residuals in the alignment
and, on occasion, by comparing the point spread functions of speciifc objects.  We
also verified the alignment after each step by visual inspection
(blinking) the positions of distributed objects within each field.

We note that for the F336W and narrow-band images, applying {\it tweakreg}
 failed initially, owing to significant contamination from cosmic rays that resulted 
 in false star matches among the individual FLT files.  For these filters, an alternate strategy was used:
 false matches were mitigated by creating a preliminary cosmic ray-cleaned image
via drizzling the pipeline FLTs (ignoring the slight misalignment of the
FLTs), constructing a custom reference catalog for this temporary image (including
careful rejection of artificial sources, especially along image edges), and then
applying {\it tweakreg} with this catalog to align the original pipeline FLTs. 

Drizzled images were constructed for each filter from the aligned
FLTs using the {\it AstroDrizzle} software. Then {\it tweakreg} was used again to align these
drizzled images to the F814W drizzled image for each field, which was selected
to be the reference image. (The only exception was for F164N
images, which were aligned instead to F160W in this step.)  The DrizzlePac 
{\it tweakback} routine was used to propagate the F814W-matched WCS solution back 
into the headers of the aligned FLTs.  Finally, all of the aligned FLTs were drizzled onto a
common footprint, producing a set of aligned images for each field.

Finally, the WCS information in the file headers of data for the individual fields was corrected 
to place all of the data on the same absolute astrometric scale, as described in the main text.

\clearpage

\begin{deluxetable}{ccccccccccc}
\rotate\tablecaption{HST WFC3 Observations of M83}
\tablehead{\colhead{Field} & 
\colhead{Item} & 
\colhead{F336W} & 
\colhead{F438W} & 
\colhead{F502N} & 
\colhead{F547M} & 
\colhead{F657N} &
\colhead{F673N} & 
\colhead{F814W} &  
\colhead{F164N} & 
 \colhead{F160W} 
  \\
\colhead{RA/Dec$^{a}$} & 
\colhead{~} & 
\colhead{U} &
\colhead{B} &
\colhead{[O~III]} &
\colhead{y} &
\colhead{H$\alpha$} &
\colhead{[S~II]} &
\colhead{I} &
\colhead{[Fe~II]} &
\colhead{H}
}
\tabletypesize{\scriptsize}
\tablewidth{0pt}\startdata
F1$^b$ &  t(s) & 1890 & 1920 & 2484 & 1203$^c$ & 1484 & 1850 & 1213 & 2397 & 2397  \\
13:37:04.43 & Date(UT) & 2009-08-26 & 2009-08-26 &  2009-08-26 &  2009-08-26 &  2009-08-26 &  2009-08-20 &  2009-08-26 &  2009-08-26 &  2009-08-26  \\
-29:51:28.00 & Flash(s) & ... & ... & ... & ... &  ... & ... & ... & na$^d$ & na$^d$  \\
F2$^b$ &  t(s) & 2560 & 1800 & 2484 & 1203 & 1484 & 1770 & 1213 & 2397 & 2397  \\ 
13:37:04.80 & Date(UT) & 2010-03-17 & 2010-03-17 & 2010-03-19 & 2010-03-20 & 2010-03-19 & 2010-03-17 & 2010-03-19 & 2010-03-17 & 2010-03-17  \\
-29:49:16.00 & Flash(s) & ... & ... & ... & ... &  ... & ... & ... & na & na  \\
F3  &  t(s) & 2579 & 1799 & 2982 & 2682 & 1799 & 2262 & 1379 & 2109 & 534  \\
13:37:06.80 & Date(UT) & 2012-08-27 & 2012-08-27 & 2012-08-27 & 2012-08-27 & 2012-08-27 & 2012-08-27 & 2012-08-27 & 2012-08-27 & 2012-08-27  \\
-29:53:55.90 & Flash(s) & 10 & 3 & 8 & 5 &  8 & 8 & None & na & na  \\
F4  &  t(s) & 2589 & 1809 & 2982 & 2682 & 1809 & 2262 & 1379 & 2109 & 534  \\
13:36:53.90 & Date(UT) & 2012-07-22 & 2012-07-22 & 2012-08-31 & 2012-08-31 & 2012-07-22 & 2012-08-31 & 2012-07-22 & 2012-08-31 & 2012-08-31  \\
-29:49:17.31 & Flash(s) & None & None & 8 & 8 &  None & 8 & None & na & na  \\
F5  &  t(s) & 2589 & 1809 & 2982 & 2682 & 1809 & 2262 & 1379 & 2109 & 534  \\
13:36:53.10 & Date(UT) & 2012-07-22 & 2012-07-22 & 2012-09-03 & 2012-09-03 & 2012-07-22 & 2012-09-03 & 2012-07-22 & 2012-09-03 & 2012-09-03  \\
-29:51:38.43 & Flash(s) & None & None & 8 & 8 &  None & 8 & None & na & na  \\
F6  &  t(s) & 2579 & 1799 & 2982 & 2682 & 1799 & 2262 & 1379 & 2109 & 534  \\
13:36:55.20 & Date(UT) & 2012-08-31 & 2012-08-31 & 2012-09-04 & 2012-09-04 & 2012-08-31 & 2012-09-04 & 2012-08-31 & 2012-09-04 & 2012-09-04  \\
-29:54:10.28 & Flash(s) & 10 & 3 & 8 & 5 &  8 & 8 & None & na & na  \\
F7  &  t(s) & 2579 & 1799 & 2982 & 2682 & 1799 & 2262 & 1379 & 2109 & 534  \\
13:37:15.74 & Date(UT) & 2012-09-04 & 2012-09-04 & 2012-09-06 & 2012-09-06 & 2012-09-04 & 2012-09-06 & 2012-09-04 & 2012-09-06 & 2012-09-06  \\
-29:51:28.00 & Flash(s) & 10 & 3 & 8 & 5 &  8 & 8 & None & na & na  \\
\tablenotetext{a}{J2000 coordinates used for the UVIS-FIX and IR-FIX pointing positions.}
\tablenotetext{b}{Fields 1 and 2 are archival observations of M83 obtained by the WFC3 SOC, program 11360.  Additional filters were used for these fields; see text.}
\tablenotetext{c}{Filter F555W was used in Field 1.}
\tablenotetext{d}{CTE Flash correction not applicable to the IR camera data.}
\enddata 
\label{table_obs}
\end{deluxetable}

\begin{deluxetable}{lccc} 
\tablecaption{Flux Conversion Factors for WFC3 Emission Line Filters}
\tablehead{\colhead{Filter} & \colhead{Conversion\tablenotemark{1}} &  \colhead{Ion} & \colhead{$\lambda$(\AA)}
}
\tablewidth{0pt}
\startdata
F502N & $3.831 \times 10^{-16}$ & [O III] & 5007 \\
F657N & $2.817 \times 10^{-16}$ & H$\alpha$+[N~II] & 6563, 6548, 6583 \\
F673N &  $3.096 \times 10^{-16}$ & [S II]  & 6716, 6731 \\
F164N &  $5.992 \times 10^{-17}$ & [Fe II]  & 16,440 
\enddata
\tablenotetext{1}{Flux in ergs $\rm cm^{-2} ~ s^{-1}$ corresponding to 1 electron per second count rate; a systemic velocity of 500 $\VEL$ for M83 was assumed to place the emission lines in the proper place in each filter bandpass.}
\label{table_conversions}
\end{deluxetable}

\begin{deluxetable}{rcccccccccl}
\rotate\tablecaption{Small Diameter and New Supernova Remnants in M83 Revealed by HST/WFC3}
\tablehead{
\colhead{} &
\colhead{} &
\colhead{} &
\colhead{} &
\colhead{} &
\colhead{} &
\colhead{GCD$^{b}$}  &
\colhead{WFC3} &
\colhead{} &
\colhead{} &
\colhead{} 
\\
\colhead{ID} &
\colhead{Prev. Name$^{a}$} &
\colhead{RA(J2000)} &
\colhead{Dec(J2000)} &
\colhead{Radius(\arcsec)} &
\colhead{Diam(pc)} &
\colhead{(kpc)} &
\colhead{Field} &
\colhead{F(F657N)$^{c}$} &
\colhead{$\frac{F(F673N)}{F(F657N)}^{d}$} &
\colhead{Comments$^{e}$}
}
\tabletypesize{\scriptsize}
\tablewidth{0pt}
\startdata
1  & B12-026	& 13:36:48.99	& $-$29:52:54.10	& 0.13	& 5.7	& 3.80	& 5	& 2.4E-16	& 0.40 &  \\
2  & B12-036	& 13:36:49.81	& $-$29:52:16.95	& 0.10	& 4.4	& 3.80	& 5	& 4.7E-16	& 0.35 &  \\
3  & ...	& 13:36:50.11	& $-$29:52:43.67	& 0.32	& 14.1	& 3.40	& 5	& $<$2.1E-15	& ... &  [Fe~II]-only; F(164N)=2.2E-16; \\
~ & ~ & ~ & ~ & ~ & ~ & ~ & ~ & ~ & ~ & \,\,\, H~II contamination. \\
4  & B12-037	& 13:36:50.12	& $-$29:53:08.78	& 0.11	& 4.8	& 3.60	& 5	& 5.5E-16	& 0.32 &  \\
5  & B12-041	& 13:36:50.55	& $-$29:53:03.88	& 0.15	& 6.6	& 3.40	& 5	& 6.2E-16	& 0.40 &  \\
6  & B12-049	& 13:36:51.02	& $-$29:53:01.33	& 0.17	& 7.5	& 3.30	& 5	& 4.8E-16	& 0.29 &  H~II contamination. \\
7  & ...	& 13:36:51.19	& $-$29:50:42.32	& 0.14	& 6.2	& 3.60	& 5	& 2.2E-15	& 0.28 &  H~II contamination. \\
8  & ...	& 13:36:51.48	& $-$29:52:33.24	& 0.56	& 24.6	& 3.00	& 5	& $<$2.5E-16	& ... &  [Fe~II] only; F(164N)=3.4E-16. \\
9  & ...	& 13:36:51.52	& $-$29:53:00.89	& 0.30	& 13.2	& 3.25	& 5	& 1.3E-14	& 0.17 &  H~II contamination. \\
10 & ...	& 13:36:51.81	& $-$29:52:01.90	& 0.23	& 10.1	& 2.75	& 5	& $<$1.2E-16	& ... &  [Fe~II] only; F(164N)=1.4E-16. \\
11 & B12-065	& 13:36:53.23	& $-$29:53:25.33	& 0.15	& 6.6	& 3.00	& 6	& 2.2E-15	& 0.24 &  H~II contamination. \\
12 & B12-067	& 13:36:53.29	& $-$29:52:48.18	& 0.24	& 10.6	& 2.50	& 5	& 1.7E-15	& 0.50 &  \\
13 & ...	& 13:36:53.73	& $-$29:48:51.26	& 0.76	& 33.4	& 5.10	& 4	& 2.1E-14	& 0.21 &  H~II contamination. \\
14 & B12-075	& 13:36:54.24	& $-$29:50:28.16	& 0.19	& 8.4	& 3.00	& 4	& 4.8E-15	& 0.45 &  \\
15 & B12-314	& 13:36:55.26	& $-$29:54:02.87	& 0.12	& 5.3	& 3.30	& 6	& 1.3E-16	& $<$0.3 &  \\
16 & B12-106	& 13:36:56.23	& $-$29:52:55.18	& 0.14	& 6.2	& 1.90	& 5	& 1.7E-15	& 0.28 &  \\
17 & B12-109	& 13:36:56.81	& $-$29:49:49.66	& 0.25	& 11.0	& 3.30	& 4	& 1.3E-15	& 0.53 &  \\
18 & B12-115	& 13:36:57.88	& $-$29:53:02.75	& 0.12	& 5.3	& 1.80	& 5	& 5.3E-15	& 0.20 &  H~II contamination. \\
19 & ...	& 13:36:58.64	& $-$29:51:06.49	& 0.09	& 4.0	& 1.43	& 5	& 4.9E-16	& 0.28 &  \\
20 & D10-02	& 13:36:58.90	& $-$29:52:26.26	& 0.29	& 12.8	& 0.90	& 1	& 1.2E-15	& 0.33 &  \\
21 & B12-119	& 13:36:59.00	& $-$29:52:56.79	& 0.08	& 3.5	& 1.50	& 5	& 6.9E-16	& 0.26 &  \\
22 & D10-03	& 13:36:59.17	& $-$29:51:47.90	& 0.13	& 5.7	& 0.60	& 1	& 4.1E-16	& 0.45 &  \\
23 & ...	& 13:36:59.32	& $-$29:48:36.51	& 0.32	& 14.1	& 4.75	& 2	& 1.5E-15	& 0.27 &  H~II contamination. \\
24 & ...	& 13:36:59.44	& $-$29:48:36.99	& 0.32	& 14.1	& 4.75	& 2	& 2.3E-15	& 0.35 &  H~II contamination. \\
25 & ...	& 13:36:59.79	& $-$29:48:37.87	& 0.30	& 13.2	& 4.75	& 2	& 5.1E-15	& 0.18 &  H~II contamination. \\
26 & B12-129	& 13:37:00.03	& $-$29:54:16.95	& 0.09	& 4.0	& 3.30	& 6	& 1.3E-16	& 0.10 &  H~II contamination. \\
27 & D10-N01	& 13:37:00.05	& $-$29:52:01.94	& 0.11	& 4.8	& 0.30	& 1	& 1.7E-15	& 0.36 &  \\
28 & D10-06	& 13:37:00.06	& $-$29:52:08.72	& 0.33	& 14.5	& 0.39	& 1	& 3.6E-15	& 0.54 &  \\
29 & D10-N04	& 13:37:00.34	& $-$29:52:05.43	& 0.14	& 6.2	& 0.28	& 1	& 4.1E-15	& 0.26 &  \\
30 & D10-N07	& 13:37:00.41	& $-$29:52:06.21	& 0.21	& 9.2	& 0.29	& 1	& 2.6E-15	& 0.50 &  \\
31 & ...	& 13:37:00.42	& $-$29:52:22.55	& 0.41	& 18.0	& 0.63	& 1	& 1.9E-16	& 1.10 &  \\
32 & ...	& 13:37:00.55	& $-$29:52:06.55	& 0.17	& 7.5	& 0.28	& 1	& 2.0E-15	& 0.38 &  \\
33 & ...	& 13:37:00.81	& $-$29:54:26.81	& 0.25	& 11.0	& 3.50	& 6	& 4.8E-15	& 0.16 & H~II contamination. \\
34 & ...	& 13:37:00.88	& $-$29:52:08.45	& 0.10	& 4.4	& 0.30	& 1	& 9.2E-16	& 0.24 &  \\
35 & B12-137=D10-09	& 13:37:01.02	& $-$29:50:56.35	& 0.25	& 11.0	& 1.41	& 1	& 1.1E-15	& 0.36 &  \\
36 & D10-10	& 13:37:01.07	& $-$29:51:41.56	& 0.24	& 10.6	& 0.34	& 1	& 1.2E-15	& 0.39 &  \\
37 & ...	& 13:37:02.00	& $-$29:51:51.42	& 0.37	& 16.3	& 0.34	& 1	& 8.8E-16	& 0.29 &  \\
38 & B12-321	& 13:37:01.28	& $-$29:51:59.89	& 0.09	& 4.0 	& 0.15	& 1	& 6.9E-16	& 0.08 &  O-strong source; see D10.\\
39 & D10-N19	& 13:37:01.61	& $-$29:52:01.91	& 0.23	& 10.1	& 0.27	& 1	& 7.3E-16	& 0.36 &  \\
40 & B12-147=BL04-40 	& 13:37:02.21	& $-$29:49:52.43	& 0.21	& 9.2	& 2.91	& 2	& 3.6E-15	& 0.39 &  \\
41 & B12-150=BL04-41	& 13:37:02.42	& $-$29:51:26.09	& 0.25	& 11.0	& 0.80	& 1	& 6.0E-15	& 0.31 &  Also D10-15.\\

42 & ...	& 13:37:02.89	& $-$29:48:39.10	& 0.66	& 29.0	& 4.64	& 2	& 1.5E-14	& 0.24 &  \\
43 & B12-151	& 13:37:03.02	& $-$29:49:45.46	& 0.18	& 7.9	& 3.08	& 2	& 5.7E-15	& 0.41 &  \\
44 & ...	& 13:37:03.14	& $-$29:54:16.92	& 0.50	& 22.0	& 3.44	& 3	& 8.2E-15	& 0.23 &  \\
45 & ...	& 13:37:03.40	& $-$29:54:02.48	& 0.53	& 23.3	& 3.25	& 3	& 8.5E-14	& 0.11 &  H~II contamination.\\
46 & B12-324=SN57D	& 13:37:03.58	& $-$29:49:40.73	& 0.09	& 4.0	& 3.22	& 2	& 1.7E-17	& 6.59 &  \\
47 & ...	& 13:37:05.02	& $-$29:55:21.73	& 0.31	& 13.6	& 5.21	& 3	& $<$1.0E-16	& ... &  [Fe~II] only; F(164N)=9.9E-16.\\
48 & ...	& 13:37:05.44	& $-$29:49:18.79	& 0.21	& 9.2	& 3.83	& 2	& 5.9E-15	& 0.06 &  H~II contamination.\\
49 & ...	& 13:37:05.88	& $-$29:50:45.46	& 0.32	& 14.1	& 2.13	& 1	& 4.5E-15	& 0.22 &  \\
50 & ...	& 13:37:06.23	& $-$29:55:05.08	& 0.17	& 7.5	& 4.82	& 3	& 2.0E-15	& 0.20 &  H~II contamination.\\
51 & D10-20	& 13:37:06.99	& $-$29:51:09.59	& 0.24	& 10.6	& 2.05	& 1	& 6.8E-16	& 0.38 &  \\
52 & B12-179	& 13:37:07.11	& $-$29:51:01.55	& 0.07	& 3.1	& 2.17	& 1	& 1.3E-15	& 0.22 &  H~II contamination.\\
53 & B12-183=D10-21	& 13:37:07.69	& $-$29:51:10.05	& 0.18	& 7.9	& 2.24	& 1	& 3.3E-15	& 0.33 &  H~II contamination. \\
54 & D10-26	& 13:37:08.33	& $-$29:50:56.36	& 0.65	& 28.6	& 2.54	& 1	& 7.1E-15	& 0.19 &  H~II contamination. \\
55 & ...	& 13:37:08.39	& $-$29:52:47.67	& 0.16	& 7.0	& 2.72	& 3	& 6.8E-15	& 0.24 &  H~II contamination. \\
56 & ...	& 13:37:08.41	& $-$29:52:10.42	& 0.51	& 22.4	& 2.37	& 1	& 5.7E-15	& 0.20 & H~II contamination.  \\
57 & B12-333=D10-29	& 13:37:08.61	& $-$29:52:42.81	& 0.44	& 19.4	& 2.7	& 1	& 1.69E-15	& 0.28 &  \\
58 & D10-35	& 13:37:09.31	& $-$29:50:58.49	& 0.57	& 25.1	& 2.79	& 1	& 1.8E-14	& 0.18 &  H~II contamination. \\
59 & B12-336	& 13:37:12.09	& $-$29:50:57.25	& 0.20	& 8.8	& 3.50	& 7	& 5.2E-15	& 0.06 &  \\
60 & ...	& 13:37:13.07	& $-$29:51:38.46	& 0.48	& 21.1	& 3.65	& 7	& 2.0E-14	& 0.30 &  H~II contamination. \\
61 & ...	& 13:37:13.33	& $-$29:51:35.98	& 0.78	& 34.3	& 3.65	& 7	& 6.7E-15	& 0.31 &  \\
62 & B12-221	& 13:37:17.20	& $-$29:51:53.37	& 0.25	& 11.0	& 5.00	& 7	& 6.2E-15	& 0.32 &  H~II contamination. \\
63 & B12-223	& 13:37:17.43	& $-$29:51:53.89	& 0.25	& 11.0	& 5.00	& 7	& 5.0E-15	& 0.42 &  H~II contamination.
\tablenotetext{a}{Previous names: B12 = Magellan catalog of Blair et al. (2012); D10 = Dopita et al. (2010). Empty entries indicate new objects from this survey.}
\tablenotetext{b}{Galactocentric distance; see text.}
\tablenotetext{c}{Units are $\rm ergs ~ cm^{-2} ~s^{-1}$. F657N includes both H$\alpha$ and [N~II] $\lambda\lambda$6548,6583.  For objects only showing [Fe II] emission, the F164N flux is given in the Comments.}
\tablenotetext{d}{Only a lower limit to the observed F([S~II])/F(H$\alpha$) ratio due to [N~II] contamination of F657N.}
\tablenotetext{e}{More extensive comments are provided in the following table.}
\enddata 
\label{table_snrlist1}
\end{deluxetable}

\begin{deluxetable}{rcccccccl}
\rotate
\tablecaption{M83 Supernova Remnant WFC3 Image Ratios and Supporting Information}
\tablehead{
\colhead{ID} &
\colhead{Prev. Name$^{a}$} &
\colhead{F(F657N)$^{b}$} &
\colhead{$\frac{F(F673N)}{F(F657N)}^{c}$} &
\colhead{$\frac{F(F164N)}{F(F657N)}^{c}$} &
\colhead{$\frac{F(F164N)}{F(F673N)}$} &
\colhead{$\frac{F(F502N)}{F(F657N)}^{c}$} &
\colhead{X-ray ID$^{d}$} &
\colhead{Comments}
}
\tabletypesize{\scriptsize}
\tablewidth{0pt}\startdata
 1 &	B12-026	& 2.4E-16	& 0.40	& ...	& ...	& 0.19	& ...	&	Compact diffuse opt neb; no [Fe~II] coverage. \\
 2 &	B12-036	& 4.7E-16	& 0.35	& 0.07	& 0.21	& 0.42	& X053	& Compact, present in all four bands; \\
 ~ & ~ & ~ & ~ & ~ & ~ & ~ & ~ & \,\,\,  H~II contam. \\
 3 &	...		& $<$2.1E-15	& $>$0.15	& $>$0.10	& 0.70	& $>$0.01	& ... & [Fe~II]-only patch in extended H~II emission. \\
 4 &	B12-037	& 5.5E-16	& 0.32	& 0.40	& 1.27	& 0.16	& X057	& [Fe~II] brt, extended knotty optical. \\
 5 &	B12-041	& 6.2E-16	& 0.40	& 0.14	& 0.35	& 0.36	& X061	& Small optical ring with [Fe~II]. \\
 6 &	B12-049	& 4.8E-16	& 0.29	& 0.11	& 0.39	& 0.31	& ...	&	Significant H~II contam. of Ft opt SNR. \\
 7 &	...		& 2.2E-15	& 0.28	& 0.28	& 1.01	& 0.14	& X067	& Brt opt knot on edge of Brt H~II; \\
  ~ & ~ & ~ & ~ & ~ & ~ & ~ & ~ & \,\,\, some H~II contamination. \\
 8 &	...		& $<$2.5E-16	& $<$0.4	& $>$1.3  & $>$3.6	& ...	& ...	&	 [Fe~II]-only patch in confused but ft \\
  ~ & ~ & ~ & ~ & ~ & ~ & ~ & ~ & \,\,\, H~II emission. \\
 9 &	...		& 1.3E-14	& 0.17	& 0.02	& 0.10	& 0.02	& ... &	Diffuse [Fe~II] patch in Vbrt H~II; \\
  ~ & ~ & ~ & ~ & ~ & ~ & ~ & ~ & \,\,\, buried SNR. \\
10 &	...		& $<$1.2E-16	& $<$0.40	& $>$1.2 & $>$2.8 & ...	& ...	&	 Isolated compact [Fe~II]; no obvious opt.\\
11 &	B12-065	& 2.2E-15	& 0.24	& 0.07	& 0.28	& 0.25	& X105	& Compact and brt, visible in all four \\
 ~ & ~ & ~ & ~ & ~ & ~ & ~ & ~ & \,\,\, bands; some H~II contam. \\
12 &	B12-067	& 1.7E-15	& 0.50	& 0.13	& 0.26	& 0.45	& X106	& Brt clumpy opt on fringes of H~II region. \\
13 &	...		& 2.1E-14	& 0.21	& 0.01	& 0.04	& 0.02	& ...	& Lg diffuse optical-[Fe~II] patch, heavily \\
 ~ & ~ & ~ & ~ & ~ & ~ & ~ & ~ & \,\,\, contam by Brt H~II. \\
14 &	B12-075	& 4.8E-15	& 0.45	& 0.16	& 0.35	& 0.16	& X121	& Brt compact opt neb in diffuse H~II; \\
 ~ & ~ & ~ & ~ & ~ & ~ & ~ & ~ & \,\,\, Vbrt more extended [Fe~II]. \\
15 &	B12-314	& 1.3E-16	& $<$0.3	& $<$0.10  & $<$0.30	 & 3.41	& (X135)	& High [O~III] neb w/possible X-ray. \\
16 &	B12-106	& 1.7E-15	& 0.28	& 1.13	& 4.10	& 0.05	& X141	& Brt compact opt w/ext bright [Fe~II]. \\
17 &	B12-109	& 1.3E-15	& 0.53	& 0.13	& 0.25	& 0.37	& X149	& Brt compact opt neb; Vbrt, more \\
 ~ & ~ & ~ & ~ & ~ & ~ & ~ & ~ & \,\,\, extended [Fe~II]. \\
18 &	B12-115	& 5.3E-15	& 0.20	& ...	& ...	& 0.27	& X159	& Brt compact opt; no [Fe~II] coverage; \\
 ~ & ~ & ~ & ~ & ~ & ~ & ~ & ~ & \,\,\, significant H~II contam. \\
19 &	...		& 4.9E-16	& 0.28	& 0.10	& 0.34	& 0.25	& X170	& Vcompact, visible in all four bands. \\
20 &	D10-02	& 1.2E-15	& 0.33	& ... & ... & 0.12	& X176	& Ft diffuse opt neb; no [Fe~II] coverage. \\
21 &	B12-119	& 6.9E-16	& 0.26	& $<$0.01 & $<$0.02 & $<$0.05	& ...	&	Compact H$\alpha$-[S~II] knot; \\
 ~ & ~ & ~ & ~ & ~ & ~ & ~ & ~ & \,\,\, significant H~II contam. \\
22 &	D10-03	& 4.1E-16	& 0.45	& ... & .... & 0.39	& X181	& Ft compact opt neb; no \\
 ~ & ~ & ~ & ~ & ~ & ~ & ~ & ~ & \,\,\, [Fe~II] coverage. \\
23 &	...		& 1.5E-15	& 0.27	& 0.14	& 0.53	& 0.06	& (X184)& Some H~II contam; strong [Fe~II].  \\
24 &	...		& 2.3E-15	& 0.35	& 0.19	& 0.54	& 0.14	& (X184)& Some H~II contam. \\
25 &	...		& 5.1E-15	& 0.18	& 0.03	& 0.16	& 0.04	& ...	& [S~II]-[Fe~II] knot in brt H~II contam. \\
26 &	B12-129	& 1.3E-16	& 0.10	& 0.07	& 0.73	& 0.73	& X199	& Ft. optical clump with X-ray source. \\ 
27 &	D10-N01	& 1.7E-15	& 0.36	& 0.02	& 0.04	& 0.20	& X202	& Compact opt neb, no [Fe~II]. \\
28 &	D10-06	& 3.6E-15	& 0.54	& 0.06	& 0.12	& 0.14	& ...	& Ext. neb visible in all four bands. \\
29 &	D10-N04	& 4.1E-15	& 0.26	& 0.08	& 0.31	& 0.20	& X212	& Compact neb visible in all four bands. \\
30 &	D10-N07	& 2.6E-15	& 0.50	& 0.21	& 0.43	& 0.03	& ...	& Compact neb with ft [Fe~II]. \\
31 &	...		& 1.9E-16	& 1.10	& 5.35	& 4.86	& $<$0.05	& X219	& [Fe~II] and X-ray source with little opt. \\
32 &	...		& 2.1E-15	& 0.38	& 0.40	& 1.06	& 0.11	& ...	& Vstr [Fe~II], larger in size \\
 ~ & ~ & ~ & ~ & ~ & ~ & ~ & ~ & \,\,\, than compact opt. neb? \\
33 &	...		& 4.8E-15	& 0.16	& 0.01	& 0.07	& 0.03	& ...	& [S~II] knot in heavy H~II contam; \\
 ~ & ~ & ~ & ~ & ~ & ~ & ~ & ~ & \,\,\, possible [Fe~II]. \\
34 &	...		& 9.2E-16	& 0.24	& 0.88	& 3.64	& 0.08	& ...	& Compact H$\alpha$-[S~II] w/Brt, more \\
 ~ & ~ & ~ & ~ & ~ & ~ & ~ & ~ & \,\,\, extended [Fe~II]. \\
35 &	B12-137=& 1.1E-15 & 0.36 & 0.10 & 0.27  & 0.43	& X235	& Ft, but visible in all four bands  \\
 ~ & D10-16 & ~ & ~ & ~ & ~ & ~ & ~ & \,\,\, plus X-ray. \\
36 &	D10-10	& 1.2E-15	& 0.39	& 0.13	& 0.35	& 0.07:	& ...	&	Vft; [Fe~II] very uncertain. \\
37 &	...		& 8.8E-16	& 0.29	& 0.30	& 1.00	& 0.09	& ...	&	Ft, but appears to have [Fe~II]. \\
38 &	B12-321	& 6.9E-16	& 0.08	& $<$0.01 & $<$0.04 & 0.98	& X243	& Compact source on eastern edge \\
 ~ & ~ & ~ & ~ & ~ & ~ & ~ & ~ & \,\,\, of Nucleus; see D10. \\
39 &	D10-N19	& 7.3E-16	& 0.36	& $<$0.01 & $<$0.01 & 0.12	& X250	& X-ray yes; [Fe~II] no; [S~II] \\
 ~ & ~ & ~ & ~ & ~ & ~ & ~ & ~ & \,\,\, is at limit for HST. \\
40 &	B12-147=	& 3.6E-15	& 0.39	& 0.40	& 1.01	& 0.19	& X261	& Vbrt and compact. \\
 ~ & BL04-40 & ~ & ~ & ~ & ~ & ~ & ~ & \,\,\, \\
41 &	B12-150= & 6.0E-15	& 0.31	& 0.09	& 0.29	& 0.21	& X265	& Brt, compact double knot; [Fe~II]. \\
~ & D10-15= & ~ & ~ & ~ & ~ & ~ & ~ & ~  \\
~ & BL04-41 & ~ & ~ & ~ & ~ & ~ & ~ & ~  \\
42 &	...		& 1.5E-14	& 0.24	& 0.01	& 0.06	& 0.03	& ...	&	Nice ring; H~II contam. \\
  ~ & ~ & ~ & ~ & ~ & ~ & ~ & ~ & \,\,\, from adj emission. \\
43 &	B12-151	& 5.7E-15	& 0.41	& 0.31	& 0.77	& 0.06	& X272	& Vbrt compact opt w/ Vbrt more \\
 ~ & ~ & ~ & ~ & ~ & ~ & ~ & ~ & \,\,\, extended [Fe~II]. \\
44 &	...		& 8.2E-15	& 0.23	& 0.04	& 0.16	& 0.03	& ...	&	Ft partial shell w/ H~II contam. \\
45 &	...		& 8.5E-14	& 0.11	& 0.01	& 0.10	& 0.04	& (X275) & Vbrt H~II with embedded SNR; \\
 ~ & ~ & ~ & ~ & ~ & ~ & ~ & ~ & \,\,\, [Fe~II] and [O~III]. \\
46 &	B12-324	& 1.7E-17	& 6.59	& $<$0.60	& $<$0.01	& 41.48	& X279	& SN1957D. \\
47 &	...		& $<$1.0E-16	& ...	& $>$10	& ... & ...	& ...	&	Vbrt [Fe~II]-only source. \\
48 &	...		& 5.9E-15	& 0.06	& $<$0.01	& $<$0.03	& 0.16	& ...	&	Compact [O~III]-[S~II] nebula in  \\
 ~ & ~ & ~ & ~ & ~ & ~ & ~ & ~ & \,\,\, brt H~II; no [Fe~II] or X-ray. \\
49 &	...		& 4.5E-15	& 0.22	& 0.16	& 0.70	& 0.06	& ...	&	Partial shell on edge of brt H~II; \\
 ~ & ~ & ~ & ~ & ~ & ~ & ~ & ~ & \,\,\, strong [Fe~II]. \\
50 &	...		& 2.0E-15	& 0.20	& 0.05	& 0.27	& 0.18	& ...	&	Ft SNR w/significant H~II contam. \\
51 &	D10-20	& 6.8E-16	& 0.38	& 0.11	& 0.29	& 0.05	& ...	&	Ft H$\alpha$-[S~II] patch w/ Ft [Fe~II]. \\
52 &	B12-179	& 1.3E-15	& 0.22	& 0.62	& 2.80	& 0.09	& (X321) & Vstr [Fe~II], more extended \\
 ~ & ~ & ~ & ~ & ~ & ~ & ~ & ~ & \,\,\, than opt; X-ray is XRB coinc. \\
53 &	B12-183= & 3.3E-15	& 0.33	& 0.07	& 0.23	& 0.11	& ...	&	Brt knot on edge of H~II; some contam. \\
~ & D10-21 & ~ & ~ & ~ & ~ & ~ & ~ & ~  \\
54 &	D10-26	& 7.1E-15	& 0.19	& 0.05	& 0.24	& 0.10	& X336	& H$\alpha$ ring in H~II contam; [Fe~II]. \\
55 &	...		& 6.79E-15	& 0.24	& 0.01	& 0.05	& 0.03	& ...	&	Linear H$\alpha$-[S~II] neb with significant \\
 ~ & ~ & ~ & ~ & ~ & ~ & ~ & ~ & \,\,\, H~II contam. \\
56 &	...		& 5.7E-15	& 0.20	& 0.07	& 0.33	& 0.02	& ...	&	Diffuse patch of [Fe~II] in confused \\
 ~ & ~ & ~ & ~ & ~ & ~ & ~ & ~ & \,\,\, H~II region. \\
57 &	B12-333= & 1.7E-15	& 0.28	& ... & ... & 0.28	& ...	&	Well-defined opt arc; no [Fe~II] coverage. \\
~ & D10-29 & ~ & ~ & ~ & ~ & ~ & ~ & ~  \\
58 &	D10-35	& 1.8E-14	& 0.18	& 0.02	& 0.14	& 0.00	& ...	&	H~II contam; [Fe~II] and [S~II] are \\
 ~ & ~ & ~ & ~ & ~ & ~ & ~ & ~ & \,\,\, clearly visible. \\
59 &	B12-336	& 5.2E-15	& 0.06	& $<$0.01 & $<$0.01 & 0.09	& X360	& Compact [S~II]-[O~III] knot \\
 ~ & ~ & ~ & ~ & ~ & ~ & ~ & ~ & \,\,\, buried in brt H~II. \\
60 &	...		& 2.0E-14	& 0.30	& 0.02	& 0.06	& 0.07	& ...	&	SNR buried in Vbrt H~II; [Fe~II] and \\
 ~ & ~ & ~ & ~ & ~ & ~ & ~ & ~ & \,\,\, [S~II] are best. \\
61 &	...		& 6.7E-15	& 0.31	& 0.05	& 0.16	& 0.07	& ...	&	Ft, extended SNR w/ Ft [Fe~II] as well; \\
 ~ & ~ & ~ & ~ & ~ & ~ & ~ & ~ & \,\,\, some contam. \\
62 &	B12-221	& 6.2E-15	& 0.32	& 0.25	& 0.79	& 0.27	& X389	& Brt compact opt w/ Vbrt ext [Fe~II]; \\
 ~ & ~ & ~ & ~ & ~ & ~ & ~ & ~ & \,\,\, some contam. \\
63 &	B12-223	& 5.1E-15	& 0.42	& 0.51	& 1.20	& 0.11	& X391	& Brt compact opt w/ Vbrt ext [Fe~II]; \\
 ~ & ~ & ~ & ~ & ~ & ~ & ~ & ~ & \,\,\, some contam. 
\tablenotetext{a}{Previous names: M = Magellan catalog (B12); D10 = Dopita et al. (2010). Empty entries indicate new objects from this survey.}
\tablenotetext{b}{Units are $\rm ergs ~ cm^{-2} ~s^{-1}$. Note: F657N includes both H$\alpha$ and [N~II] $\lambda\lambda$6548,6584.  For objects only showing [Fe II] emission, see Comments in Table 3.}
\tablenotetext{c}{Only a lower limit to the actual line ratios of interest due to [N~II] contamination in F657N filter.}
\tablenotetext{d}{X-ray ID from Chandra catalog of Long et al. (2014). Parentheses indicate less reliable alignments.}
\enddata 
\label{table_snrlist2}
\end{deluxetable}

\end{document}